\documentclass[a4paper,12pt]{article}

\usepackage{amsmath,amssymb,amsfonts}% Typical maths resource packages
\usepackage{graphicx}% Packages to allow inclusion of graphics
\usepackage{cite}
\usepackage{hyperref}% For creating hyperlinks in cross references
\usepackage{color}% For creating coloured text and background
\usepackage{caption}

\makeatletter
\@addtoreset{equation}{section}
\renewcommand{\theequation}{\thesection.\@arabic\c@equation}
\makeatother

\definecolor{red}{rgb}{1,0,0}% Standard colours red, green, blue
\definecolor{green}{rgb}{0,1,0}
\definecolor{blue}{rgb}{0,0,1}
\definecolor{darkblue}{rgb}{0,0,0.5}
\definecolor{lightblue}{rgb}{.5,.5,1}
\definecolor{lightgray}{gray}{.87}% How you can define your own greys
\definecolor{Dark}{gray}{.20}
\definecolor{pink}{rgb}{.95,0.82,0.92}% How you can define your own colours
\definecolor{yellow}{rgb}{1,1,0}
\definecolor{lightyellow}{rgb}{1,1,.5}
\definecolor{purple}{rgb}{0.7,0,0.85}
\definecolor{darkgreen}{rgb}{0,0.5,0}
\definecolor{orange}{rgb}{0.8,0.2,0.2}
\def \be {\begin{equation}}
\def \ee {\end{equation}}
\def \bea {\begin{align}}
\def \eea {\end{align}}

\def \rr {\raise.35ex\hbox{\small $\prime$}\kern-.17em{\mbox{\large $\imath$}}}
\def \del {\partial}
\def \dels {\partial\kern-.5em / \kern.5em}
\def \As {{A\kern-.5em / \kern.5em}}
\def \Ds {D\kern-.7em / \kern.5em}

\def \th {\theta}

\allowdisplaybreaks[1]

\begin{document}

\begin{titlepage}

\vspace*{-15mm}   
\baselineskip 10pt   
\begin{flushright}   
\begin{tabular}{r}
RIKEN-iTHEMS-Report-19\\    
OU-HET 1036 
% 2019
\end{tabular}   
\end{flushright}   
\baselineskip 24pt   
\vglue 10mm   

\begin{center}
{\Large\bf
An Analytic Description of \\ Semi-Classical Black-Hole Geometry
%%% PM +
%Collapsing Shell Inside Apparent Horizon with
%Back-Reaction of Vacuum Energy
}

\vspace{8mm}   

\baselineskip 18pt   

\renewcommand{\thefootnote}{\fnsymbol{footnote}}
{\large
Pei-Ming Ho${}^{a}$%
\footnote[2]{e-mail: pmho@phys.ntu.edu.tw},
Yoshinori Matsuo${}^{b}$%
\footnote[3]{e-mail: matsuo@het.phys.sci.osaka-u.ac.jp}
and 
Yuki~Yokokura$^c$%
\footnote[4]{e-mail: yuki.yokokura@riken.jp}
}
\renewcommand{\thefootnote}{\arabic{footnote}}
 
\vspace{5mm}   

{\it  
% Affiliation
${}^{a}$
Department of Physics and Center for Theoretical Physics, \\
National Taiwan University, Taipei 106, Taiwan,
R.O.C. 
\\
\vskip 3mm
${}^{b}$
Department of Physics, Osaka University, \\
Toyonaka, Osaka 560-0043,
Japan
 \\
\vskip 3mm
${}^{c}$
iTHEMS Program, RIKEN, Wako, Saitama 351-0198, Japan
}
  
\vspace{10mm}   

\end{center}

\begin{abstract}
%%% 0406-1
%%% Just rewriting the first two sentences into one sentence:
We study analytically the spacetime geometry
of the black-hole formation and evaporation. 
As a simplest model of the collapse, we consider a spherical thin shell, 
and take the back-reaction from the negative energy of the quantum vacuum state. 
%%% The following statement is not entirely new.
%Due to the back-reaction of the negative energy of the quantum vacuum state, 
%the black-hole geometry has a wormhole-like structure around the Schwarzschild radius. 
%In the dynamical process, 
%the neck of the wormhole-like structure becomes the apparent horizon. 
%%% We have also analyzed the geometry outside the horizon.
%In this paper, 
%we investigate the geometry inside the apparent horizon. 
%%% This is an important specific feature of our paper
For definiteness, we will focus on quantum effects of s-waves.
%%% Main result:
We obtain an analytic solution of the semi-classical Einstein equation
for this model, that provides an overall description of 
the black hole geometry form the formation to evaporation.
%%% 2nd main result:
As an application of this result,
we find its interesting implication that,
after the collapsing shell enters the apparent horizon,
the proper distance between the shell and 
the horizon remains as small as the Planck length
even when the difference in their areal radii 
is of the same order as the Schwarzschild radius. 
The position of the shell would be regarded as the same place 
to the apparent horizon in the semi-classical regime of gravity. 
%as long as the Schwarzschild radius is sufficiently larger than the Planck scale. 
%until the black hole is a small fraction of its initial mass.
%%% 0406-2
\end{abstract}

\baselineskip 18pt   

\end{titlepage}

\newpage

\baselineskip 18pt

\noindent\rule{\textwidth}{1pt}

\setcounter{tocdepth}{2}
\tableofcontents

\vskip 12pt

\noindent\rule{\textwidth}{1pt}

\vskip 12pt

%\clearpage

%%%%%%%%%%%%%%%%%%%%%%%%%%%%%%%%%%%%%%%%%%%%%%%%%%%%%%%%%%%%%%%%%%%%%%%%%%%%%%%%
%%%%%%%%%%%%%%%%%%%%%%%%%%%%%%%%%%%%%%%%%%%%%%%%%%%%%%%%%%%%%%%%%%%%%%%%%%%%%%%%
%%%%%%%%%%%%%%%%%%%%%%%%%%%%%%%%%%%%%%%%%%%%%%%%%%%%%%%%%%%%%%%%%%%%%%%%%%%%%%%%

\section{Introduction}

% YM: I would like to avoid to use the word ``conventional model'' 
% since only you knows what it means. 
% Also, it is not a good idea to say that 
% the purpose of this paper is to study the conventional model. 
% We should say that we are studying a physical black hole but with some approximation. 

%%% PM +
%[Try to be logically continuous.]
%YM-11/12
Quantum effects around the black holes have been
well studied in the literature.
% YM: I think DFU is too detailed to put in the basic concept of the paper. 
%It is a scenario of gravitational collapses
%including the quantum effect of the vacuum energy-momentum tensor
%following the ideas of Refs.\cite{Davies:1976ei,Christensen:1977jc}.
%YM-11/12
%%%PM: It is not a salient feature as it is not true for the KMY model.
%A salient feature of the effects is that
%YM-12/16
% YM: I think ``Many believe...'' implicitly means ``we do not agree with it.''
%Many believe that a common feature of the quantum effects 
% YM: I wrote the following assuming that ``common...in many models'' is common only in the models.
A common feature of the quantum effects in many models is that 
there is an incoming negative vacuum energy flux
%-12/16
around the apparent horizon \cite{Davies:1976ei,Christensen:1977jc},
%%%MP
and the outgoing positive vacuum energy flux
(i.e. Hawking radiation)
appears well outside the apparent horizon.
In this scenario,
when the collapsing matter falls under the apparent horizon,
there is no known mechanism to transfer 
all information of the collapsing matter into Hawking radiation
when the black-hole mass is still macroscopic,
unless there are high-energy events around the apparent horizon
\cite{Mathur:2009hf,firewall}.

The goal of this paper is to have a better understanding of
the spacetime geometry for black holes,
in particular for the region inside the apparent horizon.
We study the quantum effect of the vacuum energy-momentum tensor 
used in Refs.\cite{Davies:1976ei,Christensen:1977jc}, 
and take its back-reaction to the geometry into account. 
While there are works of numerical simulation,
e.g.\ Ref.~\cite{Parentani:1994ij},
we take the analytic approach
to see detailed features of the geometry.
%%% 0406-1
%YM-4/8
As we will see,
the analytic solution will allow us to learn 
detailed feature which is difficult to be found in numerical simulations.
%-4/8
%%% 0406-2
Throughout this paper,
we focus on spherically symmetric configurations for simplicity.

%%% 0406-1
An intuitive picture of the near-horizon geometry
for a dynamical black hole is already known in the literature.
%%% 0406-2
With the back-reaction of quantum effects taken into consideration, 
the black-hole geometry outside the collapsing matter
has a wormhole-like structure near the horizon 
due to the negative vacuum energy
\cite{wormhole-as-BH,Fabbri,Ho:2017joh,Ho:2017vgi,Berthiere:2017tms,Ho:2018fwq}.
For black holes with spherical symmetry,
it corresponds to a local minimum of the areal radius
which we will refer to as the ``neck'' of the near-horizon geometry.
The areal radius $a$ of the neck is approximately 
the Schwarzschild radius associated with the total mass of the collapsing matter. 
In the dynamical process of the evaporation,
the neck is shrinking with time, 
and plays the role of the apparent horizon in the geometry. 
This picture has been clearly demonstrated in numerical simulation \cite{Parentani:1994ij}
as well as analytic calculation \cite{Ho:2018jkm}.

%%% 0406-1
In this work,
we aim at finding explicit analytic solutions to 
the semi-classical Einstein equation
that are more explicit than existing results in the literature,
with which one may explore new features of the black-hole geometry.
%%% Sentences are reshuffled but unchanged in the following two paragraphs.
%%% 0406-2
%YM-4/8 
% Before we mention about the shell
% We have to write explicitly that we will consider a collapsing shell as a simplest model. 
As a simple example of the gravitational collapse,
we consider a collapsing thin shell with the spherical symmetry. 
%-4/8
We shall provide explicit descriptions of 
the whole spacetime geometry,
from the interior of the collapsing shell to large distances. 
We describe the geometry after the shell enters the apparent horizon, 
until the black hole is evaporated to a microscopic scale 
where the low-energy effective theory breaks down. 

\begin{figure}
\begin{center}
 \includegraphics[scale=0.3]{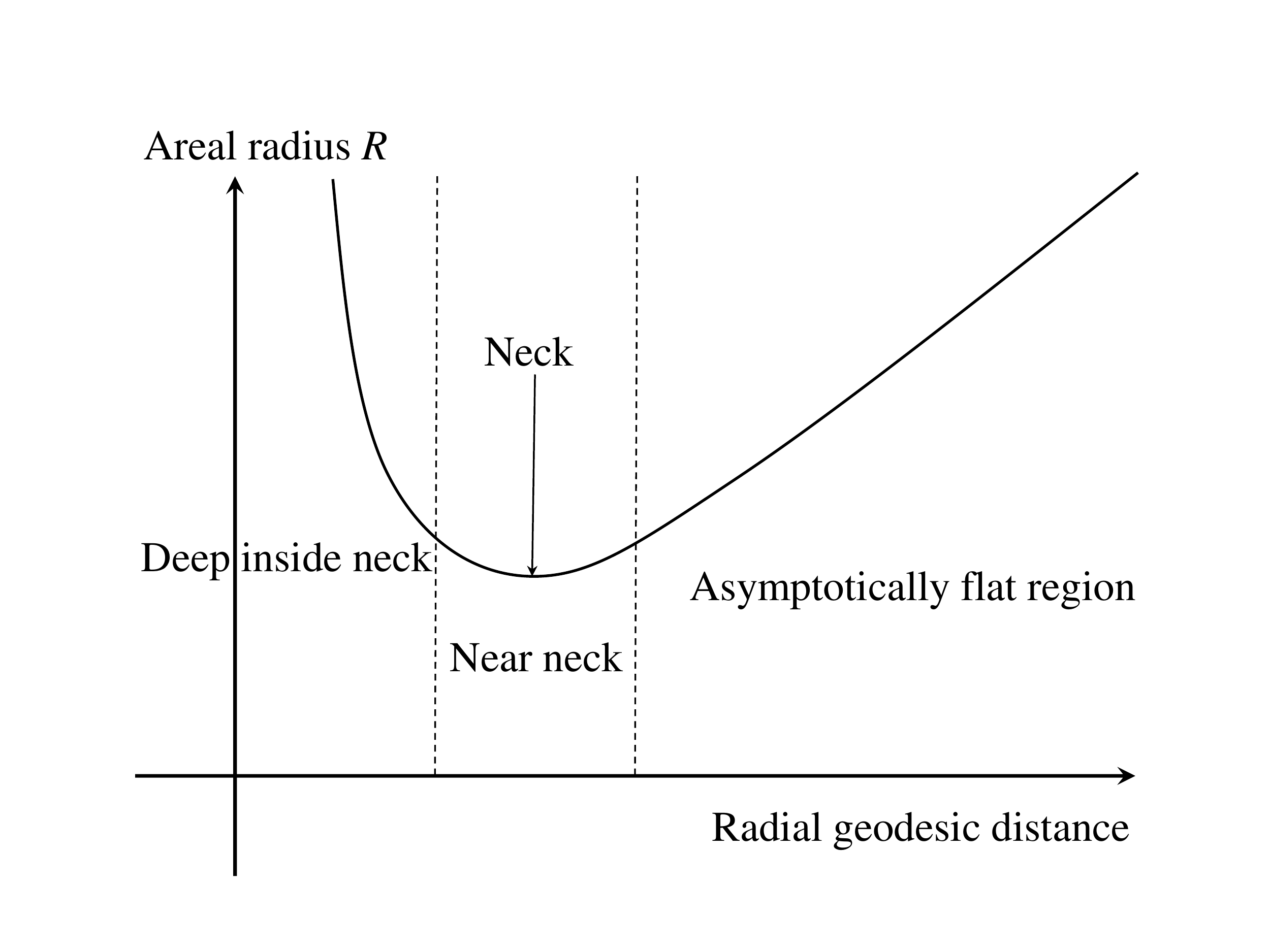}
 \caption{\small
 The areal radius $R$ and 3 regions, asymptotically flat, near the neck and deep inside the neck. 
 The areal radius has the local minimum, which is the apparent horizon in the dynamical case. 
 The geometry near the neck is studied in \cite{Ho:2018jkm}, 
 and the relation between its time evolution and the junction condition at the shell 
 are discussed in Sec.~\ref{sec:neck-shell}. 
 The geometry outside the near-neck region is referred to as the asymptotically flat region. 
 The geometry in this region will be analyzed by using the perturbative expansion around 
 the outgoing Vaidya metric in Sec.~\ref{sec:vaidya-pert}. 
 The areal radius becomes larger also inside the neck. 
 We call this region as the region deep inside the neck 
 and investigate the geometry in this region in Sec.~\ref{sec:deep-dyn}. 
 }\label{fig:regions}
\end{center}
\end{figure}

%YM-4/8 
The geometry around the collapsing shell is obtained 
by connecting the interior and exterior geometries at the shell. 
The interior geometry is simply given by the flat space 
as it is not affected by the collapsing shell. 
The exterior geometry is the black hole geometry with back reaction, 
which is the semi-classical version of the (exterior) Schwarzschild metric and 
contains the near-horizon geometry explained above. 
We divide the exterior geometry into 3 regions:
%-4/8 
the asymptotically flat region, 
the region near the neck, 
and the region deep inside the neck (See Fig.~\ref{fig:regions}).
The semi-classical Einstein equation can be solved 
in different regions by different perturbative expansions. 
The initial condition is imposed in the past before the gravitational collapse, 
and the boundary conditions for each region
are given by the junction conditions.%

%YM-4/8 
%%%% 0406-1
%As our goal is to achieve rigorous, explicit solutions 
%to the semi-classical Einstein equation for a dynamical geometry,
%this paper includes a lot of technical details.
%Furthermore, 
%for completeness and cross-checking,
%our analysis is done in different coordinate systems
%with our results scattered in Secs.\ref{sec:neck-shell}--\ref{sec:deep-dyn}.
%The lengthy detailed calculation is justified in the end
%by its intriguing implication
%that the near-horizon geometry has a Planckian nature unnoticed before,
%which is worthy of future investigation.
%%%% 0406-2
%-4/8 

%%% 0406-1
Our plan for this paper is as follows.
%%% 0406-2
In Sec.~\ref{sec:review},
we review the static geometry for a black hole. 
The vacuum state, which is known as the Boulware vacuum,
has no divergence because the black-hole geometry 
is modified from the classical solution 
by the back-reaction from the negative vacuum energy. 
The exterior geometry of an evaporating black hole 
can then be obtained by introducing the Hawking radiation to 
the static black-hole geometry. 
In Sec.~\ref{sec:neck-shell},
after reviewing the dynamical geometry outside the neck \cite{Ho:2018jkm},
we study the time evolution of the geometry and that of the shell. 
In Sec.~\ref{sec:vaidya-pert},
we consider the perturbative expansion 
in the asymptotically flat region and explore the junction condition 
to the region near the neck.
The result suggests that the difference between the areal radius
at the collapsing shell and that at the apparent horizon 
becomes large as the shell moves to the deeper region, 
but their proper distance remains as small as the Planck length. 
To further verify this result,
we investigate the region deep inside the neck in Sec.~\ref{sec:deep-dyn}. 
The result is consistent with Sec.~\ref{sec:vaidya-pert}. 
Sec.~\ref{sec:conclusion} is devoted to the conclusion and discussions.

%%%%%%%%%%%%%%%%%%%%%%%%%%%%%%%%%%%%%%%%%%%%%%%%%%%%%%%%%%%%%%%%%%%%%%%%%%%%%%%%
%%%%%%%%%%%%%%%%%%%%%%%%%%%%%%%%%%%%%%%%%%%%%%%%%%%%%%%%%%%%%%%%%%%%%%%%%%%%%%%%
%%%%%%%%%%%%%%%%%%%%%%%%%%%%%%%%%%%%%%%%%%%%%%%%%%%%%%%%%%%%%%%%%%%%%%%%%%%%%%%%

\section{Static Black-Hole Geometry}
\label{sec:review}

%YM-11/12
% YM: In the preamble of Sec.2, we just introduce the energy-momentum tensor of DFU. 
% YM: The following paragraph is moved to Sec.2.1. 
%%%% PM +
%%[The following paragraph is moved here from the Introduction.]
%%%% PM -
%The vacuum state for the static geometry,
%which is known as the Boulware vacuum,
%has neither incoming nor outgoing radiation at large distances.
%It is well-known that the vacuum energy-momentum tensor
%of the Boulware vacuum diverges at the horizon in a first-order perturbative calculation
%for the Schwarzschild background.
%But, in fact, no divergence appears if the back-reaction of the vacuum energy to geometry
%is properly taken into account \cite{Ho:2017joh}. 
%Furthermore,
%it is possible to introduce the time-dependence of 
%the dynamical geometry for the Unruh vacuum
%as a perturbation to this non-singular static geometry \cite{Ho:2018jkm}.
%%%% PM +
%\footnote{
%The Unruh vacuum can be treated as a perturbation of the Boulware vacuum
%only in suitable coordinate systems.
%}
%As part of this work will be based on such a perturbation theory,
%we review the static geometry for the Boulware vacuum here.
%%%% PM -
%%YM-10/30
%-11/12

This section is a review of previous results included in
Refs.\cite{Ho:2017joh,Ho:2017vgi,Ho:2018jkm},
where the s-wave approximation is used for the matter fields 
and the vacuum energy-momentum tensor is assumed to
be given by the toy model proposed in Refs.\cite{Davies:1976hi,Davies:1976ei}
based on 2-dimensional massless scalar fields.%
%YM-10/10
\footnote{%
The most general 4D static, spherically symmetric black-hole solutions
with vacuum energy-momentum tensor of 4D conformal matters 
are studied in Ref.\cite{Ho:2018fwq}. 
%PM-10/05
% I do not understand the following sentence, so I remove it.
%It is argued that only incoming and outgoing modes give 
%significant quantum effects, and back-reaction from 
%the vacuum energy-momentum tensor around the Schwarzschild radius 
%would be independent of the details of matter fields, 
%but determined by the boundary condition for those modes, 
%namely, $q$ in Ref.\cite{Ho:2018fwq}. 
The toy model studied here is a special case
belonging to the class of models with $q < 0$ in Ref.\cite{Ho:2018fwq}. 
%YM-11/12
%as a simple example of the conventional model of black holes.
%-11/12
}
%-10/10
(The solution was presented in different coordinate systems in
Ref.~\cite{Ho:2017joh,Ho:2017vgi},
and here we use the same coordinate system as Ref.\cite{Ho:2018jkm}.)
%YM-11/12
As a preliminary of the review, 
we introduce the vacuum energy-momentum tensor 
in Refs.~\cite{Davies:1976hi,Davies:1976ei} first. 

%In this paper, 
%-11/12
We consider only spherically symmetric configurations 
in 4D spacetime.
The metric can always be written as
\begin{equation}
 ds^2 = - C(u,v) du\,dv + R^2(u,v) d \Omega^2 \ ,
 \label{metric0}
\end{equation}
where $u$ and $v$ are retarded and advanced (null) time coordinates 
which span the 2-dimensional subspace of the temporal and radial directions, 
and $d \Omega^2$ is the metric on a unit 2-sphere, 
\begin{equation}
 d \Omega^2 \equiv d \theta^2 + \sin^2 \theta d\phi^2 \ .  
\end{equation}
We refer to the radius $R(u, v)$ of the symmetric 2-spheres as the {\em areal radius}.
%since the radial distance from the origin is in general different from $R(u,v)$,
%while the area of a symmetric 2-sphere (with fixed values of $u, v$, centered at the origin)
%is always given by $4\pi R^2(u, v)$. 

%YM-11/12
%In this work,
%-11/12
We assume that the semi-classical Einstein equation
\begin{equation}
 G_{\mu\nu} = \kappa \langle T_{\mu\nu} \rangle
\end{equation}
determines the spacetime geometry to a good approximation.
%$G_{\mu\nu}$ is the classical Einstein tensor and 
The constant $\kappa$ is related to the Newton constant $G_N$ via $\kappa = 8 \pi G_N$.
Note that the quantum effect is taken into consideration 
through the vacuum expectation value $\langle T_{\mu\nu} \rangle$
of the energy-momentum tensor.

As mentioned above,
we choose $\langle T_{\mu\nu} \rangle$
to be given by that of the 2D scalar fields \cite{Davies:1976hi,Davies:1976ei}
as a concrete model.
In vacuum,
it is fixed by the conservation law and the Weyl anomaly
up to integration constants $\beta(u)$ and $\gamma(v)$ as
\footnote{
For simplicity,
we have chosen the normalization factor $N$ to be $24\pi$
so that $\alpha = \kappa$ in the notation of Ref.\cite{Ho:2018jkm}.
}
\cite{Davies:1976hi,Davies:1976ei}
\begin{align}
\langle T_{uu} \rangle &=
- \frac{2}{R^2} C^{1/2} \partial_u^2 C^{-1/2} + \frac{\beta(u)}{R^2} \ , 
\label{Tuu-vac}
\\
\langle T_{vv} \rangle &=
- \frac{2}{R^2} C^{1/2} \partial_v^2 C^{-1/2} + \frac{\gamma(v)}{R^2} \ , 
\label{Tvv-vac}
\\
\langle T_{uv} \rangle &=
- \frac{1}{R^2 C^2}
\left[ C \partial_u\partial_v C - \partial_u C \partial_v C \right],
\label{Tuv-vac}
\\
\langle T_{\theta\theta} \rangle &= 0 \ ,
\label{Tthth-vac}
\end{align}
where $C$ and $R$ are functions defined in the metric \eqref{metric0}.

At large distances where
\begin{equation}
 r - a \gg \frac{\kappa}{a} \ ,
\end{equation}
$\langle T_{\mu\nu} \rangle$ should be very small,
and the metric is approximated by the Schwarzschild metric, 
\begin{align}
 ds^2 &= %- f(r) dt^2 + \frac{dr^2}{f(r)} + r^2 d \Omega^2 \ , 
 - \left(1 - \frac{a}{r}\right) dt^2 + \frac{dr^2}{1 - \frac{a}{r}} + r^2 d \Omega^2 \ , 
%\\
% f(r) &= 1 - \frac{a}{r} \ , 
\end{align}
where $a$ is the Schwarzschild radius. 
However, the perturbative correction is in general not weak
around the Schwarzschild radius when
\cite{Ho:2018fwq}
\be
r - a \sim \mathcal{O}\left(\frac{\kappa}{a}\right) \ . 
\ee

%%%%%%%%%%%%%%%%%%%%%%%%%%%%%%%%%%%%%%%%%%%%%%%%%%%%%%%%%%%%%%%%%%%%%%%%%%%%%%%%
%%%%%%%%%%%%%%%%%%%%%%%%%%%%%%%%%%%%%%%%%%%%%%%%%%%%%%%%%%%%%%%%%%%%%%%%%%%%%%%%

\subsection{Geometry Around The Neck}
\label{sec:neck-stat}

%YM-11/12
%%%% PM +
%%[The following paragraph is moved here from the Introduction.]
%%%% PM -
The vacuum state for the static geometry,
which is known as the Boulware vacuum,
has neither incoming nor outgoing radiation at large distances. 
It is given by the condition 
\begin{align}
 \beta(u) &= 0 \ , 
 &
 \gamma(v) &= 0
 \label{static-beta-gamma}
\end{align}
in the vacuum energy-momentum tensor \eqref{Tuu-vac}-\eqref{Tthth-vac}.
It is well-known that the vacuum energy-momentum tensor
of the Boulware vacuum diverges at the horizon in the perturbative calculation
around the Schwarzschild background.
But, in fact,
no divergence appears if the back-reaction of the vacuum energy
is properly taken into account \cite{Ho:2017joh}. 
Furthermore, 
it is possible to introduce the time-dependence of 
the dynamical geometry for the Unruh vacuum ($\beta(u) \neq 0$)
as a perturbation to this non-singular static geometry \cite{Ho:2018jkm}.%
%%% PM +
\footnote{
The Unruh vacuum can be treated as a perturbation of the Boulware vacuum
only in suitable coordinate systems.
}
As part of this work will be based on such a perturbation theory,
%we review the static geometry for the Boulware vacuum here.
%%%%% PM -
%%%YM-10/30
%
%
%As we will construct the dynamical geometry outside of the collapsing matter 
%from static geometry of the black hole, 
%by introducing the effect of the Hawking radiation $\beta(u)$ as perturbation, 
we review the static geometry near the Schwarzschild radius
\cite{Ho:2017joh,Ho:2017vgi,Ho:2018jkm}.

Here, we consider the geometry near the Schwarzschild radius 
with the back-reaction from the vacuum energy-momentum tensor. 
%The vacuum energy-momentum tensor is given by \eqref{Tuu-vac}-\eqref{Tthth-vac}. 
%We consider the Boulware vacuum, 
%\begin{align}
% \beta(u) &= 0 \ , 
% &
% \gamma(v) &= 0 \ .  
% \label{static-beta-gamma}
%\end{align}
%Although the quantum effect diverges at 
%the Schwarzschild radius in $\kappa$-expansion around the Schwarzschild metric, 
%the geometry becomes regular if we take the back-reaction into account. 
%%as we will briefly review \cite{us...}.
%
%-11/12
For static metrics, it is convenient to use the time coordinate 
associated with the timelike Killing vector. 
Since the areal radius $R$ may not be a single-valued function in the radial direction, 
it is better to take a different radial coordinate. 
In terms of the tortoise coordinate $x$ for the radial direction, 
the metric eq.\eqref{metric0} is expressed as 
\begin{equation}
 ds^2 = - C(x) \left(dt^2 - dx^2 \right) + R^2(x) d \Omega^2 \ . 
 \label{static-ansatz}
\end{equation}
At large distances,
we can consider the $\kappa$-expansion around the Schwarzschild solution
\begin{equation}
 C \simeq 1 - \frac{a}{R} \ .
\end{equation}
In the near-horizon region where
the areal radius $R$ is very close to the Schwarzschild radius,
$R - a \sim \mathcal O(\kappa/a)$, 
the Schwarzschild solution becomes comparable to its quantum correction. 

We should define the $\kappa$-expansion
in the near-horizon region as
\begin{align}
 C(x) &= \kappa C_0(x) + \mathcal O(\kappa^2/a^4) \ , 
 \\
 R(x) &= a + \kappa R_0(x) + \mathcal O(\kappa^2/a^3) \ . 
\end{align}
In the following,
from time to time,
$\mathcal{O}(\kappa^n/a^m)$
will be denoted as $\mathcal{O}(\kappa^n)$ for simplicity,
as the power $m$ of $a$ is uniquely determined by the dimension of the quantity.

The solution to the semi-classical Einstein equations to the 1st order has
\cite{Ho:2017joh,Ho:2017vgi,Ho:2018jkm}
\begin{align}
 C_0 &= c_0 e^{k x} \ , 
 \label{c-neck-stat}
\\
 R_0 &= a_1 + \frac{c_0}{a k^2} e^{k x} - \frac{k}{4 a} x \ .
 \label{r-neck-stat}
\end{align}
To determine the parameters $k$, $c_0$ and $a_1$,
we consider the continuation of this solution to the region
where the $\kappa$-expansion around the Schwarzschild solution is valid.
\footnote{
In Ref.\cite{Ho:2018jkm},
these parameters are not determined from the junction condition.
Instead, they are fixed by assuming the relation $\dot a(u) = - \kappa \beta(u)$
between the time evolution of the Schwarzschild radius $a(u)$
and the Hawking radiation $\beta(u)$.
}

It was pointed out in Ref.\cite{Ho:2018fwq} that,
in terms of the coordinate $z$ defined by
\begin{equation}
 ds^2 = - C dt^2 + dz^2 + R^2 d \Omega^2 \ ,
\end{equation}
the perturbative expansion around the Schwarzschild solution
can be extended to the near-horizon region
where $R - a \sim \mathcal O(\kappa/a)$.
%and give a consistent result to the expansion for $R - a \sim \mathcal O(\kappa/a)$ above, 
%
%The Schwarzschild metric is expanded around the Schwarzschild radius as 
%\begin{align}
% C &= \frac{z^2}{4 a^2} + \cdots \ , 
%\\
% R^2 &= a^2 + \frac{1}{2} z^2 + \cdots \ ,  
% \label{r-z-neck-stat-1st}
%\end{align}
%The linear order correction in $\kappa$ expansion for $R^2$ is calculated as 
%\begin{equation}
% \kappa a_2 - \kappa \log \frac{z}{a} \ , 
%\end{equation}
%where $a_2$ is an integration constant. 
%These correction terms are comparable to non-trivial part of the radius, $R^2-a^2$, 
%in the Schwarzschild solution \eqref{r-z-neck-stat-1st} for $z^2 \sim \mathcal O(\kappa)$. 
%Thus, $\kappa$-expansion around the Schwarzschild metric implies 
%that the metric is approximated near the Schwarzschild radius as 
The expansion to the first-order correction gives \cite{Ho:2018fwq}
\begin{align}
 C &= \frac{z^2}{4 a^2} + \cdots \ , 
\\
 R^2 &= a^2 + \frac{1}{2} z^2 + \kappa a_2 - \kappa \log \frac{z}{a} + \cdots \ , 
\end{align}
near the Schwarzschild radius.
Although the first-order correction to the areal radius $R$ 
is comparable to the 0th-order term (given by the Schwarzschild solution) at small $z$,
no higher-order correction is of a comparable order of magnitude \cite{Ho:2018fwq}.
Hence,
after the coordinate transformation to the tortoise coordinate, 
the solution above reproduces eqs.\eqref{c-neck-stat} and \eqref{r-neck-stat}. 

In this way,
the parameters in eqs.\eqref{c-neck-stat} and \eqref{r-neck-stat} are fixed
such that
\begin{align}
 C &= \frac{1}{4} e^{(x-x_0)/a} + \cdots \ , 
 \label{c-x-pert-stat}
\\
 R^2 &= a^2 + \frac{a^2}{2} e^{(x-x_0)/a} 
  - \frac{\kappa}{2a} (x-x_0)
  %+ \kappa a_2 + \kappa \log 2
  + \cdots \ ,
 \label{r-x-pert-stat}
\end{align}
where the constant part of the 1st-order terms in $R^2$
is absorbed by a redefinition of $a$.
(Without loss of generality, 
one can set $x_0$ to $0$.)
%The integration constant is determined as 
%\begin{align}
% k &= \frac{1}{a} \ , 
% &
% c_0 &= \frac{1}{4\kappa} e^{-x_0/a}  \ , 
% &
% a_1 &= \frac{x_0}{4 a^2} + \frac{a_2}{2 a} + \frac{1}{2 a} \log 2 \ . 
%\end{align}
%Although the expansion around the Schwarzschild solution 
%is valid only for $z^2 \gg \mathcal O(\kappa)$, 
%it gives a good approximation even around $z^2 \sim \mathcal O(\kappa)$ 
%if the metric is expressed in terms of the proper radial coordinate $z$. 

The result above implies that the areal radius $R$ has a local minimum at
\begin{align}
% z_0 \simeq \sqrt{\kappa}
x_\text{neck} = x_0 - a \log\left(\frac{a^2}{\kappa}\right) \ ,
\label{xneck}
\end{align}
where
\begin{align}
 R^2(x_\text{neck}) &= a^2
 % + \kappa a_2 
  + \frac{\kappa}{2}\left(\log \frac{a^2}{\kappa} +1\right) + \mathcal O(\kappa^2) \ .
 %& \text{at} \qquad
 %z = z_0,
\end{align}
In the following,
we will refer to this local minimum of $R$ as the ``neck''.

Around the neck,
there is no event horizon 
and the vacuum energy-momentum tensor has no divergence.
However, eqs.\eqref{c-x-pert-stat} and \eqref{r-x-pert-stat}
provide a good approximation of the metric only 
in the near-horizon region where $R - a \sim \mathcal O(\kappa)$. 
In the next subsection,
we consider the geometry deeper inside the neck.

%%%%%%%%%%%%%%%%%%%%%%%%%%%%%%%%%%%%%%%%%%%%%%%%%%%%%%%%%%%%%%%%%%%%%%%%%%%%%%%%
%%%%%%%%%%%%%%%%%%%%%%%%%%%%%%%%%%%%%%%%%%%%%%%%%%%%%%%%%%%%%%%%%%%%%%%%%%%%%%%%

\subsection{Geometry Deep Inside The Neck}
\label{sec:deep-stat}

In the previous subsection,
we reviewed the static spacetime geometry in vacuum around the neck of a black hole.
Here we turn to the region deep inside the neck (but still in vacuum)
\cite{Ho:2017vgi},
assuming that the surface of the star is
further deeper inside the neck.

Since the areal radius $R$ is a local minimum at the neck,
it becomes larger as we go deeper inside the neck. 
When $R$ is close to the Schwarzschild radius
so that $R - a \sim \mathcal O(\kappa)$, 
the solution \eqref{c-x-pert-stat}-\eqref{r-x-pert-stat}
is a good approximation. 
In this subsection,
we consider the geometry deeper inside the neck
where 
\begin{align}
R -a \sim \mathcal O(\kappa^0) \ . 
\end{align}
Although we refer to this region as ``deep inside the neck'',
it does not imply that the proper distance in the radial direction
between the neck and a point in this region is much larger than the Planck length.
In fact,
we will see in Sec.\ref{sec:distance-deep} that it is of the order of the Planck length.

If we naively extend the solution \eqref{c-x-pert-stat}-\eqref{r-x-pert-stat} 
for $R - a \sim \mathcal O(\kappa)$ to the deeper region
where $R -a \sim \mathcal O(\kappa^0)$, 
we deduce from eq.\eqref{r-x-pert-stat} that
\begin{align}
|x - x_0| \sim \mathcal O(\kappa^{-1})
\label{x-x0}
\end{align}
(note that $x - x_0 < 0$ inside the neck),
so that $\log C \sim \mathcal O(\kappa^{-1})$. 
Thus, we expect that $C$ behaves as $\log C \sim \mathcal O(\kappa^{-1})$ 
for $R -a \sim \mathcal O(\kappa^0)$. 
Defining $\rho$ by
\begin{equation}
 C = e^{2\rho} \ , 
 \label{rho-def}
\end{equation}
we have the expansions
\begin{align}
 \rho &= \kappa^{-1} \rho_0 + \rho_1 + \mathcal O(\kappa) \ , 
\\
 R &= R_0 + \kappa R_1 + \mathcal O(\kappa^2) \ ,
\end{align}
deep inside the neck.

The solutions to the leading-order terms of the semi-classical Einstein equation are found to be
%\begin{align}
% 0 &= \partial_x \left(\rho_0 + R_0^2\right) \ , 
%\\
% 0 &= 2 \partial_x^2\rho_0 + \partial_x^2 R_0^2  \ . 
%\end{align}
%The solution of these differential equations is calculated as 
\begin{align}
 \rho_0 &= - c_1 (x-x_0) + c_2 \ , 
 \label{rho-x-deep-stat}
\\
 R_0^2 &= c_1 (x-x_0) + c_3 \ , 
 \label{r-x-deep-stat}
\end{align}
where $c_1$, $c_2$ and $c_3$ are integration constants. 

%Although the perturbative expansion around the Schwarzschild solution, 
%\eqref{c-x-pert-stat}-\eqref{r-x-pert-stat} is invalid in this region, 
%it is still consistent with the result above. 
%For $x - x_0\sim \mathcal O(\kappa^{-1})$ with $x-x_0 < 0$, 
%the second term in \eqref{r-x-pert-stat} is negligible and 
%\eqref{c-x-pert-stat}-\eqref{r-x-pert-stat} 
%agrees with \eqref{rho-x-deep-stat}-\eqref{r-x-deep-stat}. 
%The integration constants are identified as 
%YM-10/10
The result above implies that \eqref{c-x-pert-stat}--\eqref{r-x-pert-stat} 
give a good approximation at the leading order even in this region. 
%-10/10
The patching of this solution
with eqs.\eqref{c-x-pert-stat}-\eqref{r-x-pert-stat} 
around the neck demands that
\begin{align}
 c_1 &\simeq - \frac{\kappa}{2 a} \ , 
& 
 c_2 &\simeq 0 \ ,
 %c_1 x_0 \ , 
& 
 c_3 &\simeq a^2 \ . 
\end{align}
It may appear strange that $c_1$ is of the 1st order in $\kappa$,
while, by definition, $R_0$ should be of the 0th order.
This is simply a result of eq.\eqref{x-x0} in this region.
%It should be noted that the conditions for the integration constant $c_1$ is not good 
%since we are considering $\kappa$-expansion assuming that 
%all the integration constants are of $\mathcal O(\kappa^0)$. 
%The constant $c_1$ becomes of $\mathcal O(\kappa)$ because 
%the tortoise coordinate $x$ is not suitable to describe the solution in this region, 
%and behaves as $|x - x_0| \sim \mathcal O(\kappa^{-1})$, which is also inconsistent with 
%the assumption that the coordinates are of $\mathcal O(\kappa^0)$. 
%In order to solve this problem, we can just introduce a new coordinate as 
In terms of a properly rescaled coordinate such as
$\tilde x = \kappa (x-x_0)$, 
all the parameters and coordinates are of $\mathcal O(\kappa^0)$.%
%YM-10/10
\footnote{%
One may say that $x$ is not a good coordinate to describe the solution in this region,
and $\tilde x$ is more appropriate,
but the results are equivalent. 
} 
%-10/10

%%%%%%%%%%%%%%%%%%%%%%%%%%%%%%%%%%%%%%%%%%%%%%%%%%%%%%%%%%%%%%%%%%%%%%%%%%%%%%%%
%%%%%%%%%%%%%%%%%%%%%%%%%%%%%%%%%%%%%%%%%%%%%%%%%%%%%%%%%%%%%%%%%%%%%%%%%%%%%%%%
%%%%%%%%%%%%%%%%%%%%%%%%%%%%%%%%%%%%%%%%%%%%%%%%%%%%%%%%%%%%%%%%%%%%%%%%%%%%%%%%

\section{Collapsing Shell}
\label{sec:neck-shell}

The geometry around the neck for a static black hole
in the Boulware vacuum
was discussed above in Sec.\ref{sec:neck-stat}.
In this section,
we study the dynamical geometry around the neck
for a black hole with the back-reaction from
the vacuum energy-momentum tensor for the Unruh vacuum. 
%YM-11/12
The geometry is obtained by introducing the Hawking radiation as perturbation 
to the static case of Sec.~\ref{sec:neck-stat}. 
%-11/12
As a simple model for the collapsing matter,
we consider a thin shell falling at the speed of light. 
The matter distribution in the radial direction is approximated by 
the Dirac delta function.
%and the shell is located on a null hypersurface. 
\footnote{
An infinitesimally thin shell introduces unphysical solutions
\cite{Ho:2019qpo} but they are avoided in our perturbative approach.
}
%Although the metric is given in the form of eq.\eqref{metric0} 
%both inside and outside of the shell, 
%it is more convenient to consider separately
%the metric inside the shell and the metric outside,
%and demand that they satisfy the junction conditions. 
%
%The metric outside the shell is in the form of eq.\eqref{metric0},
%and the metric inside the shell is simply that of the flat spacetime. 
%The geometry outside the shell is curved by the collapsing shell, 
%and the energy-momnetum tensor receives non-trivial quantum corrections. 

We will study the dynamical geometry
outside the collapsing matter \cite{Ho:2018jkm}. 
%YM-10/10
The geometry for a collapsing thin shell can be obtained by 
connecting this exterior geometry with the interior geometry 
(which is flat spacetime) at the collapsing shell. 
We will first consider the generic amplitude $\beta(u)$ of Hawking radiation,
%-10/10
and then determine the Hawking radiation by
imposing the junction condition on the collapsing shell.

\subsection{Geometry Outside Collapsing Shell}

%YM-10/10
%For a black hole formed by the gravitational collapse of a thin shell,
%the geometry inside the shell is the flat spacetime,
%-10/10
%while the geometry outside the shell is curved due to the collapsing shell, 
%as well as non-trivial quantum corrections. 
The geometry outside the shell is a solution to
the semi-classical Einstein equation \eqref{Tvv-vac}--\eqref{Tthth-vac}. 
%The integration constants $\beta(u)$ and $\gamma(v)$
%in eqs. \eqref{Tuu-vac} and \eqref{Tvv-vac}
%are determined by the junction condition at the collapsing shell 
%and the initial condition in the past null infinity. 
%Assuming the vacuum state in the past null infinity, 
%we take 
%\begin{equation}
% \gamma(v) = 0 \ . 
%\end{equation}
%The junction condition for $\beta(u)$ depends on physics of the shell. 
We impose the initial condition that there are no incoming matter excitations 
in the past null infinity other than the collapsing null shell. 
By identifying the null coordinates $u$ and $v$ with those 
in the asymptotically flat spacetime with $C \rightarrow 1$, 
the integration constant $\gamma(v)$
in the incoming energy flow $\langle T_{vv}\rangle$ \eqref{Tvv-vac}
should vanish:
\begin{equation}
 \gamma(v) = 0  \ ,
 \label{gamma=0}
\end{equation}
as we assumed \eqref{static-beta-gamma} for the static solution.
The other integration constant $\beta(u)$
in the outgoing energy flow $\langle T_{uu}\rangle$ \eqref{Tuu-vac}
is identified with the Hawking radiation
and should be determined by the junction condition across the collapsing matter. 
%If the collapsing matter can be approximated by a thin shell, 
%we can consider the junction condition to the flat spacetime on the shell. 

%%%PM
%In this paper, we treat the Hawking radiation $\beta(u)$
%as a perturbation to the static geometry. 
%The static case is then given by the condition $\beta(u) = 0$, 
%which corresponds to the Boulware vacuum. 
%As we will review below, the divergence in the Boulware vacuum does not appear 
%if the back-reaction from the vacuum energy-momentum tensor
%$\langle T_{\mu\nu} \rangle$ \eqref{Tuu-vac}-\eqref{Tthth-vac} are taken into account. 
%Naively, 
%one may consider treating the dynamical geometry as
%a perturbation of the static solution.
%However,
%the energy-momentum tensor of the Boulware vacuum
%is of the Planck scale at the horizon,
%while it is weak for the Unruh vacuum.
%Hence a naive perturbative expansion starting with the static solution may not work.
%On the other hand,
As the $u$-dependence of the dynamical solution
is turned on by the non-zero $\beta(u)$, 
which contributes to the semi-classical Einstein equation at $\mathcal O(\kappa)$,
%in powers of $\del_u$ for $\beta(u)\sim \mathcal O(\kappa^0)$, 
the $u$-dependence of the solution must be very weak
so that we can consider the derivative expansion in powers of $\del_u$, 
with $\partial_u \sim \mathcal O(\kappa)$.% 
\footnote{
More precisely,
in the $(t, x)$ coordinate system,
the static solution \eqref{static-ansatz} depends on $x$ but not on $t$.
Clearly,
the same solution can be written in the $(u, x)$ coordinate system
with no dependence on $u$.
When $\beta(u)$ is turned on at $\mathcal{O}(\kappa^0)$,
the semi-classical Einstein equation implies
a $u$-dependence with $\partial_u \sim \mathcal{O}(\kappa)$.
This is not true in the $(u, v)$ coordinate system.
Whenever we say $\partial_u \sim \mathcal{O}(\kappa)$,
the $u$-derivative is taken with a fixed radial coordinate.
}
%The solution gets $u$-dependence in the integration constants 
%and the time evolution is also determined by the semi-classical Einstein equation.
%%% Copied from above (end)
%
%YM-11/12
% YM: The following paragraph is removed since it may lead to misunderstanding 
% YM: that the expansion is around a some moment of time. 
%Assuming that $\beta(u)$ is not very large but of order $\mathcal O(\kappa^0)$,
%the total energy of the black hole is changed by a fraction of
%$\mathcal{O}(\kappa/a^2)$ over a time scale of $\mathcal{O}(a)$,
%during which the black hole is to a good approximation a static configuration.
%Like an adiabatic process,
%the metric can be approximated by a similar expansion 
%as in Sec.~\ref{sec:neck-stat} near the Schwarzschild radius, 
%but with additional $u$-dependence.
%
%YM-10/10

Now, the semi-classical Einstein equation can be solved by using $\kappa$-expansion. 
As it can be seen in the explicit form of the expansion, 
the corrections from the effects of $u$-dependence appear only from the next-to-leading order, 
since it always comes with $u$-derivative, $\partial_u \sim \mathcal{O}(\kappa)$.   
%-11/12
Thus, the solution of the semi-classical Einstein equation %the metric \eqref{metric0} 
at the leading order looks the same as those in the static case, 
\eqref{c-x-pert-stat} and \eqref{r-x-pert-stat}, 
but the integration constants are now $u$-dependent functions,%
\footnote{%
Although the integration constants depend on $u$, 
their $u$-derivatives are $\mathcal O(\kappa)$. 
}
%-10/10
\begin{align}
 C(u,v) 
 &= 
 \frac{1}{4} e^{(x-x_0(u))/a(u)} + \mathcal O(\kappa^2) \ , 
 \label{C-neck}
\\
 R(u,v) &= 
 a(u) + \kappa a_1(u) + \frac{a(u)}{4} e^{(x-x_0(u))/a(u)} 
  - \frac{\kappa}{4 a(u)^2} \left(x - x_0(u)\right) + \mathcal O(\kappa^2) \ , 
 \label{R-neck}
\end{align}
where $x$ is now viewed as a function of $u$ and $v$:
\begin{equation}
 x = \frac{1}{2}\left(v - u\right) + x_0(u) \ . 
 \label{x2uv}
\end{equation}
The expressions \eqref{C-neck} and \eqref{R-neck} are valid under the assumptions
\be
C = \mathcal O(\kappa),
\qquad
x - x_A(u) = \mathcal O(a) \ . 
\label{near-horizon-region}
\ee
It should be noted that the derivative expansion is valid at each moment of $u$, 
and hence eqs.\eqref{C-neck} and \eqref{R-neck} are valid for arbitrary $u$ 
as long as eq.\eqref{near-horizon-region} is satisfied.
%$a^2(u) \gg \kappa$. 

The integration constant $a_1(u)$ can be absorbed by a redefinition of the Schwarzschild radius $a(u)$
and hence we can set it to $0$. 
The reference point $x_0(u)$ of the $x$-coordinate 
can be absorbed by a coordinate transformation of $u$.
But we shall choose $u$ to agree with the $u$-coordinate in the asymptotically flat region, 
%YM-11/12
so $x_0(u)$ will not be set to zero. 
Note that the assumption $\partial_u \sim \mathcal O(\kappa)$ 
now implies that $\dot a(u) = \mathcal O(\kappa)$ and $\dot x_0(u) = \mathcal O(\kappa)$, 
where the dots on $\dot{a}$ and $\dot x_0$ stand for the $u$-derivative. 
The junction condition is consistent with these conditions
as we will see below.
%-11/12

%YM-10/10
The semi-classical Einstein equation
gives constraints on the boundary condition on a Cauchy surface.
They are the differential equations for the integration constants, $a(u)$ and $x_0(u)$.  
At the leading order,
we have 
\begin{equation}
 \dot a(u) = - \kappa \beta(u) + \mathcal O(\kappa^2) \ . 
 \label{da/du-neck}
\end{equation}
%where the dot of $\dot{a}$ stands for the $u$-derivative. 
%YM-11/12
Thus, the condition $\dot a(u) = \mathcal O(\kappa)$ is consistent 
with the junction condition if $\beta(u) = \mathcal O(\kappa^0)$. 
The other integration constant $x_0(u)$ should be fixed 
by the coordinate patch to the asymptotically flat region for the $u$-coordinate. 
%-11/12

%YM-10/10
The areal radius has a local minimum on constant-$u$ slices at 
\begin{align}
 x_A(u) = x_0(u) - a(u) \log\left(\frac{a^2(u)}{\kappa}\right) \ ,
\label{xapp}
\end{align}
which coincides with eq.\eqref{xneck},
except that the parameters depend on $u$ here. 
Since the areal radius is always decreasing with $u$, along constant-$v$ lines, 
$x=x_A(u)$ is the trapping horizon (the trajectory of the apparent horizon).

%YM-10/24
%%% PM +
%[The following paragraph is removed because the same things are written elsewhere.]
%It should be noted that the assumption $\partial_u \sim \mathcal O(\kappa)$ 
%implies that the $u$-dependence of all integration constants are very small. 
%In eqs.\eqref{C-neck} and \eqref{R-neck},
%we have used the same identification 
%of the integration constants to that in the static case. 
%This would be reasonable since the junction condition at each moment of time 
%can be applied if the time evolution of the geometry is sufficiently slow. 
%However, the integration constant $x_0$ is not a physical parameter, 
%but only shifts the definition of the coordinate $x$. 
%It can absorbed by the redefinition of $u$ even for the dynamical case, \eqref{x2uv}. 
%At the same time, $x_0(u)$ describes the time evolution of the reference point $x=x_0$, 
%and hence is related to the structure of the geometry. 
%This implies that the time evolution of points in $x$ 
%can be absorbed by the coordinate transformation if we focus only on the geometry near the neck. 
%However, we cannot use the coordinate transformation once 
%we identify the $u$-coordinate near the neck with that in the asymptotic region. 
%Then, $x_0(u)$ will be important to see the time evolution of the geometry, 
%although we cannot determine it only from the junction condition in the static case. 
%We will discuss this problem, later. 
%-10/24
%%% PM -

%%%%%%%%%%%%%%%%%%%%%%%%%%%%%%%%%%%%%%%%%%%%%%%%%%%%%%%%%%%%%%%%%%%%%%%%%%%%%%%%
%%%%%%%%%%%%%%%%%%%%%%%%%%%%%%%%%%%%%%%%%%%%%%%%%%%%%%%%%%%%%%%%%%%%%%%%%%%%%%%%

\subsection{Energy Conservation on Collapsing Shell}

%We first consider the energy-momentum tensor for the collapsing shell. 
%We assume that the energy-momentum tensor of the shell also can be 
%interpreted as that of the 2D scalar field, 
%or equivalently, that the shell have no pressure in the angular directions, 
%$\langle T_{\theta \theta}\rangle = 0$. 
%It is convenient to introduce the 2D energy-momentum tensor $T_{\mu\nu}^{2D}$ as 
%\begin{equation}
% \langle T_{\mu\nu} \rangle = \frac{1}{R^2} T_{\mu\nu}^{2D} \ . 
%\end{equation}
%Then the incoming mode of the energy-momentum tensor 
%$T_{vv}^{2D}$ satisfies the conservation equation, 
%\begin{equation}
% \partial_u T_{vv}^{2D} = - C \partial_v \left(C^{-1} T_{uv}^{2D}\right) \ , 
% \label{Cons-Tvv}
%\end{equation}
%where the trace part of the energy-momentum tensor $T_{uv}^{2D}$ is 
%completely determined by the Weyl anomaly in terms of the 2-dimensional curvature. 
For the dynamical geometry outside the collapsing shell
given by eqs.\eqref{C-neck}--\eqref{R-neck},
$\langle T_{uv}\rangle$ \eqref{Tuv-vac} is calculated as%
%YM-10/10
\footnote{%
We need to take the higher-order corrections
studied in Ref.\cite{Ho:2018jkm} into consideration 
to derive this expression. 
} 
%-10/10
\begin{equation}
% T_{uv} = - \frac{C}{2 a^2(u)} + \mathcal O(\kappa^2) \ . 
 \langle T_{uv}\rangle = - \frac{C(u, v)}{2 a^2(u) R^2(u, v)} + \mathcal O(\kappa^2) \ . 
\end{equation}
while 
\begin{equation}
 \langle T_{uv}\rangle = 0 \ , 
\end{equation}
in the flat spacetime inside the collapsing shell. 

Without loss of generality,
we will choose the collapsing shell to be located at $v = 0$,
and it passes through the point $x = x_0$ at $u = 0$.
The energy-momentum tensor $\langle T_{uv}\rangle$
is expressed in terms of the step function $\Theta$ as 
\begin{equation}
 \langle T_{uv}\rangle = - \frac{C}{2 a^2(u) R^2(u, v)} \Theta(v) + \mathcal O(\kappa^2) \ . 
 \label{Tuv-vac-2}
\end{equation}
The conservation law %\eqref{Cons-Tvv}
\begin{equation}
 \partial_u (R^2 \langle T_{vv}\rangle) = - C \partial_v \left(C^{-1} R^2 \langle T_{uv}\rangle\right) \ , 
 \label{Cons-Tvv}
\end{equation}
gives 
\begin{equation}
 \partial_u \langle T_{vv}\rangle = \frac{C}{2a(u)^2 R^2(u, v)} \delta(v) + \cdots \ , 
 \label{Tvv-shell}
\end{equation}
where we only keep the term proportional to the delta function. 
Eq.~\eqref{Tvv-shell} implies that the energy on the collapsing shell $\langle T_{vv}\rangle$ 
is increasing, as the right-hand side of eq.\eqref{Tvv-shell} is positive,
despite the fact that the total mass of the black hole is decreasing. 
The increase of the energy on the shell comes from
the vacuum energy sitting on top of the shell, 
and the total mass of the black hole decreases
due to the negative vacuum energy inflow.

The other component of the conservation equation is
\begin{equation}
 \partial_v (R^2 \langle T_{uu}\rangle)= - C \partial_u \left(C^{-1} R^2 \langle T_{uv}\rangle\right) \ .  
 \label{Cons-Tuu}
\end{equation}
If the collapsing shell directly loses energy into the Hawking radiation,
$\langle T_{uu}\rangle$ must have a discontinuity across the shell. 
The equation above implies that a contribution
proportional to the delta function in $\langle T_{uv}\rangle$ 
is necessary for the discontinuity of $\langle T_{uu}\rangle$. 
As the delta function is absent in
the vacuum contribution to $\langle T_{uv}\rangle$ \eqref{Tuv-vac-2},
the only chance for $\langle T_{uu}\rangle$ to be discontinuous is that
the collapsing shell contributes a delta function term in $\langle T_{uv}\rangle$.
Since the shell is collapsing at the speed of light, or equivalently, 
it lies on a null surface,
the only component the shell can contribute to is $\langle T_{vv}\rangle$. 
Therefore,
$\langle T_{uu}\rangle$ has to be continuous on the collapsing shell. 
\footnote{
This is a consequence of the assumption of no pressure,
i.e. $\langle T_{\th\th}\rangle = \langle T_{\phi\phi}\rangle = 0$.
}
Since the outgoing energy is zero
in the flat spacetime inside the collapsing shell
($\langle T_{uu}\rangle = 0$ at $v = 0^-$), 
we must have
\be
\langle T_{uu}\rangle = 0
\ee
just outside of the collapsing shell at $v = 0^+$. 
In the next subsection,
we will use this condition to determine
the magnitude of Hawking radiation.

%%%%%%%%%%%%%%%%%%%%%%%%%%%%%%%%%%%%%%%%%%%%%%%%%%%%%%%%%%%%%%%%%%%%%%%%%%%%%%%%
%%%%%%%%%%%%%%%%%%%%%%%%%%%%%%%%%%%%%%%%%%%%%%%%%%%%%%%%%%%%%%%%%%%%%%%%%%%%%%%%

\subsection{Locus of Collapsing Shell}

%YM-10/10
As the areal radius of the shell $R_s$ is given by eq.\eqref{R-neck}, % at $v = 0$.
its time-derivative is expressed as 
\begin{equation}
 \dot R_s(u) = 
 \dot a(u)
 %- \frac{1}{8} e^{(x-x_0(u))/a(u)} 
  - \frac{1}{8} e^{-\frac{u-u_0(u)}{2a(u)}} 
  + \frac{\kappa}{8 a(u)^2} + \mathcal O(\kappa^2) \ ,
  \label{dRs/du-neck}
\end{equation}
where we have chosen the position of the shell as $v = 0$, 
and $u_0$ comes from the redefinition of $u$. 
%is defined as $u_0(u) = v_0 - 2 x_0(u)$.
%-10/10

%We assume that the energy-momentum tensor of the shell
%is given by the s-wave approximation,
%which we have introduced for calculation of the vacuum energy-momentum tensor, 
%and the tangential pressure on the shell vanishes identically, 
%\begin{equation}
% T_{\theta\theta}^\text{shell} = 0 \ . 
%\end{equation}
%Then the conservation law in 2D spacetime in temporal and radial directions 
%implies that the outgoing energy flow $T_{uu}$ must be continuous at the shell. 
%%The outgoing energy flow $T_{uu}$ must be zero 
%%in the flat space inside the collapsing shell. 
%%Then, the integration constant $\beta(u)$ must be chosen such that 
%%$\langle T_{uu}\rangle = 0$ on (the vicinity of) the collapsing shell. 

The time evolution of the Schwarzschild radius $\dot a(u)$ is related to 
the outgoing radiation via eq.\eqref{da/du-neck}. 
According to eq.\eqref{Tuu-vac},
the outgoing energy flow $\langle T_{uu}\rangle$ 
at the neck is
\begin{align}
 \langle T_{uu} \rangle 
% &= 
% \frac{1}{R^2} \left(- 2 C^{1/2} \partial_u^2 C^{-1/2} + \beta(u) \right)
%\notag\\
 &= 
 \frac{1}{R^2} \left( - \frac{1}{8 a^2(u)} + \beta(u) + \mathcal O(\kappa) \right) \ . 
\end{align}
The continuity condition of the outgoing energy flow across the collapsing shell
fixes $\beta(u)$ as 
\begin{equation}
 \beta(u) = \frac{1}{8 a^2(u)} \ , 
 \label{Hawking-neck}
\end{equation}
so that eqs.\eqref{da/du-neck} and \eqref{dRs/du-neck} give
\begin{equation}
\dot{a} = - \frac{\kappa}{8a^2} + \mathcal{O}(\kappa^2),
\label{da/du-neck-2}
\end{equation}
and 
\begin{equation}
 \dot R_s = 
% - \frac{1}{8} e^{(x-x_0(u))/a(u)} 
 - \frac{1}{8} e^{-\frac{u - u_0(u)}{2a(u)}} 
 + \mathcal O(\kappa^2) \ . 
 \label{dRs/du-neck-2}
\end{equation}
%YM-10/10

The radius of the shell continues to decrease even after passing the neck
(which is the local minimum of $R$ with respect to the variation in $x$).
This is in fact necessary for the continuity of $R$ across the shell, 
as the areal radius of the incoming null surface must be monotonically decreasing 
from the viewpoint of an observer in the flat spacetime inside the shell.
In terms of the $U$ coordinate in the flat spacetime, 
$R_s$ behaves as 
\begin{equation}
 \partial_U R_s = - \frac{1}{2} \ ,
\end{equation}
which is consistent with the junction condition
\begin{equation}
 \frac{dU}{du} = C = \frac{1}{4} e^{(x-x_0(u))/a(u)} \ .
\end{equation}
%it is consistent with the locus of the shell near the neck, \eqref{dRs/du-neck}. 
%\begin{equation}
% \partial_u R_s = \frac{dU}{du} \partial_U r = - \frac{1}{2} C = - \frac{1}{8} e^{x/a(u)} \ .  
%\end{equation}
%As the shell is moving to a deeper place ($x \ll x_0$)
%inside the Schwarzschild radius $a$
%where the radius $R$ is much larger than $a$.
%YM-10/10
%On the other hand,
It can be seen by comparing eq.\eqref{dRs/du-neck-2} with eq.\eqref{da/du-neck-2}
that since $R_s(u)$ decreases at a rate that becomes exponentially small,
the Schwarzschild radius $a(u)$ will decrease much faster than $R_s(u)$.
Thus the difference $R_s(u)-a(u)$
is increasing while the radius $R_s(u)$ itself is decreasing.

\begin{figure}
\begin{center}
 \includegraphics[scale=0.4]{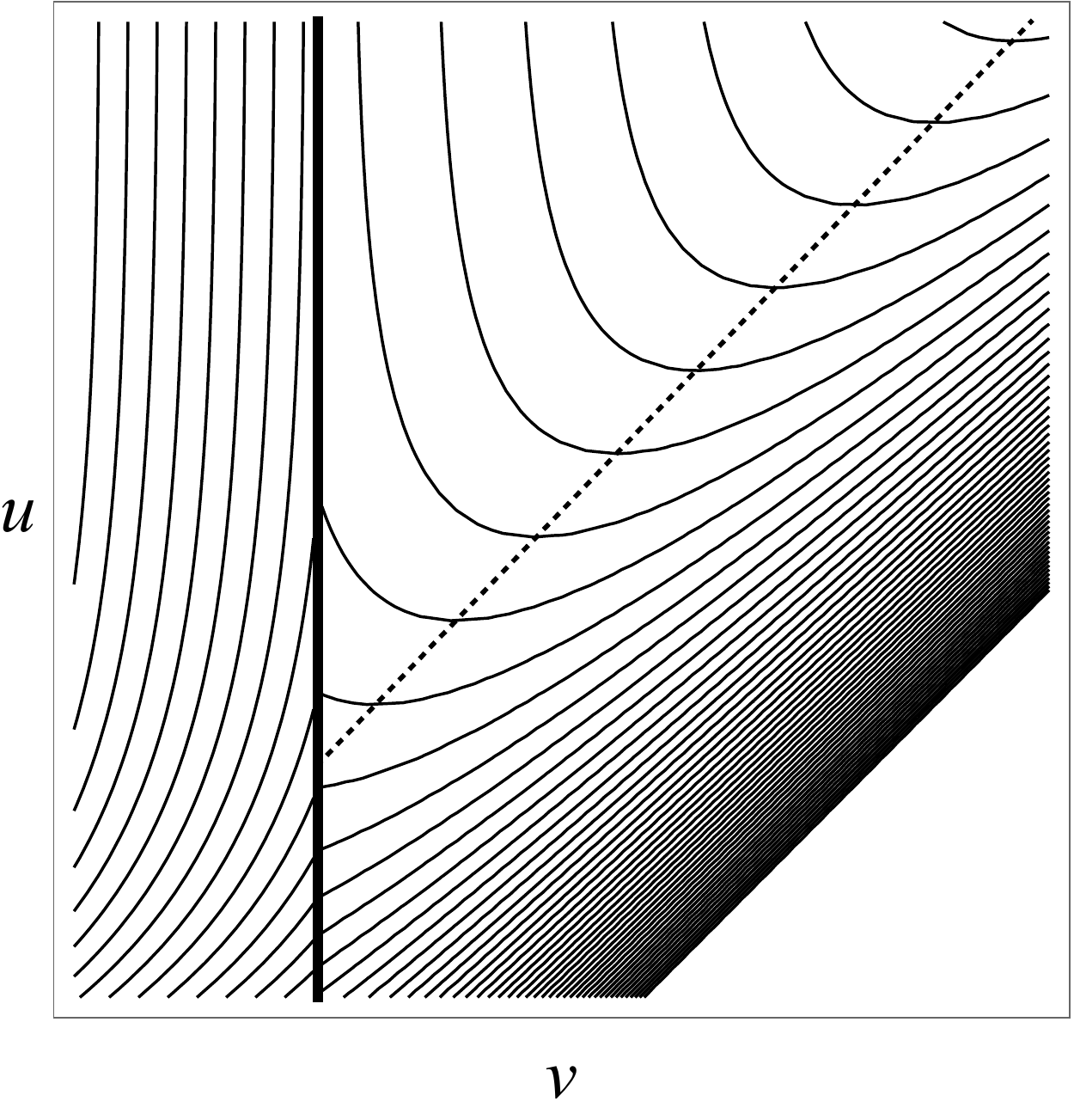}
 \caption{\small
 Lines of $R(u,v)=\text{const.}$ in the geometry near the neck. 
 Lines are just omitted in the lower-right of the figure since there are too many lines. 
 The bold line is the collapsing thin shell. 
 The dotted line is the neck, or equivalently, the apparent horizon. 
 Lines on the upper side has smaller $R$ than those on the lower side since $\frac{\partial R}{\partial u} < 0$. 
 The lines on the right has larger $R$ than those on the left, namely $\frac{\partial R}{\partial v} > 0$, 
 except for the trapped region between the collapsing shell and the apparent horizon, 
 where $\frac{\partial R}{\partial v} < 0$. 
 }\label{fig:contour}
\end{center}
\end{figure}

%YM-4/8
In this section, we studied the geometry near the neck, 
the junction condition at the shell and their time evolution. 
The geometry is given by connecting the exterior geometry \eqref{C-neck} and \eqref{R-neck} 
with flat interior spacetime at the collapsing shell, $v=0$. 
The Schwarzschild radius $a(u)$ has the time evolution according to \eqref{da/du-neck-2}. 
The structure of the geometry is shown in Fig.~\ref{fig:contour}. 
The geometry is consistent with that by Parentani and Piran, \cite{Parentani:1994ij}. 
Thus the geometry of the evaporating black hole is interpreted as 
the time evolution of a space which has the structure of the ``Wheeler's bag of gold.''

It should be noted that our result in this section relies 
on the assumption that the time evolution of $x_0(u)$ is sufficiently small, $\dot x_0(u) = \mathcal O(\kappa)$. 
%-4/8
%%% PM +
According to eq.\eqref{dRs/du-neck-2},
the decrease in $R_s$ from $u_1$ to $u_2$ is
\be
\int_{u_1}^{u_2} du \, \dot{R}_s \simeq \frac{a}{4}\left(e^{-\frac{u_1-u0}{2a}}-e^{\frac{u_2-u_0}{2a}}\right),
\ee
which is bounded from above by $a/4$ as long as $u_2 > u_1 > u_0$.
This implies that the areal radius at the collapsing shell is still $\mathcal O(a)$
(while its decrease can be as large as $a/4$)
when the shell is still in the neighborhood \eqref{near-horizon-region}
where the approximation is good.
%%% PM -
It should however be noted that the time evolution of $x_0(u)$ is assumed to be slow; 
$\dot x_0(u) = \mathcal O(\kappa)$. 
If instead $\dot x_0(u) = \mathcal O(\kappa^0)$,
eq.\eqref{dRs/du-neck-2} becomes 
\begin{equation}
 \dot R_s 
 = 
 - \frac{1}{8} e^{-\frac{u}{2a(u)}} \left( 1 + 2 \dot x_0(u) \right) 
 + \frac{\kappa \dot x_0(u)}{4 a^2(u)} + \mathcal O(\kappa^2) \ . 
\end{equation}
Thus, the discussion about $R_s(u)$ above depends on the assumption on $x_0(u)$. 
%Though it is always possible to remove $x_0(u)$ by using the coordinate transformation of $u$, 
%we also assume that it becomes $u = t-r$ in the flat spacetime in the asymptotic region. 
In order to check whether $\dot x_0(u) = \mathcal O(\kappa)$, 
we should study the junction condition in the dynamical case in more details. 
In the next section,
we consider the perturbative expansion around the outgoing Vaidya metric, 
which is valid for $r - a = \mathcal O(a)$ outside the apparent horizon, 
and show that the areal radius at the collapsing shell is still large
as long as the shell is still in the neighborhood of the neck \eqref{near-horizon-region}.

%%%%%%%%%%%%%%%%%%%%%%%%%%%%%%%%%%%%%%%%%%%%%%%%%%%%%%%%%%%%%%%%%%%%%%%%%%%%%%%%
%%%%%%%%%%%%%%%%%%%%%%%%%%%%%%%%%%%%%%%%%%%%%%%%%%%%%%%%%%%%%%%%%%%%%%%%%%%%%%%%
%%%%%%%%%%%%%%%%%%%%%%%%%%%%%%%%%%%%%%%%%%%%%%%%%%%%%%%%%%%%%%%%%%%%%%%%%%%%%%%%

\section{Back-Reaction to Vaidya Metric}
\label{sec:vaidya-pert}

%%% PM +
In the previous section,
we have studied the black-hole geometry near the neck.
In this section,
we aim at connecting the geometry near the neck
with the geometry in the asymptotically flat region.

To do so,
%%% PM -
we study the spacetime geometry 
in a different coordinate system,
for which the most general spherically symmetric metric in 4D spacetime is of the form
\begin{equation}
 ds^2 = - f(u,r) du^2 - 2 du\,dr + R^2(u,r) d \Omega^2 \ , 
 \label{vaidya}
\end{equation}
where we have imposed the gauge-fixing condition $g_{ur} = g_{ru} = -1$.
%This is a generalization of the outgoing Vaidya metric.
%YM: This is the most general spherically symmetric metric. 
% It is not very good to call it as Vaidya metric. 
%YM-11/12
The geometry is assumed to be asymptotically flat, 
$f(u,r) \to 1$ and $R(u,r) \to r$ 
at the spatial infinity $r\to \infty$. 
%-11/12
%In the previous section, we only focused on the geometry near the neck. 
%In order to see the whole structure of the spacetime, 
%Perturbative expansion around the classical solution would be convenient 
%since it is good except for the near horizon region or deeper. 

The coordinates in \eqref{vaidya} is convenient for the outgoing Vaidya metric, 
which describes the geometry of the evaporating black hole around the asymptotic region, 
since only the Hawking radiation would be important in the quantum effects there. 
In order to solve the semi-classical Einstein equation for the black-hole geometry, 
we will consider the perturbation theory of the metric \eqref{vaidya} 
starting with the outgoing Vaidya metric. 
Although the areal radius $R$ is commonly identified as the radial coordinate $r$, 
by imposing the gauge condition $R=r$ instead of $g_{ur} = g_{ru} = -1$, 
it is not very useful to study the back-reaction to the Vaidya metric, 
because the areal radius $R$ 
has a local minimum and is thus not single-valued if the back-reaction is taken into account. 
This is why we adopt another radial coordinate $r$ in the metric \eqref{vaidya}.
As above, we assume that $\partial_u \sim \mathcal O(\kappa)$ in the $\kappa$-expansion.

%YM-12/16
% YM: We may remove this paragraph, since it is not necessary to summarize the result of Sec.4. 
% This paragraph is to explain the purpose of Sec.4.3, 
% which is just to check the validity of the result near the neck. 
%-12/16
%
%YM-11/12
%As we will see, the higher-order corrections become comparable 
%to the leading-order terms if $r - a = \mathcal O(\kappa)$, 
%and hence, the perturbation around the Vaidya metric does not work well. 
%{\color{blue} 
%(What is the logical relation between the first clause and the 2nd clause
%in the next sentence?)
%}
As the expansion around the outgoing Vaidya metric is formally valid 
%YM-12/16
only 
%-12/16
for $r - a = \mathcal O(a)$ outside the Schwarzschild radius, 
we will consider another perturbation for $r - a = \mathcal O(\kappa)$
%YM-12/16
in order to check the validity of the result from the expansion around the outgoing Vaidya metric. 
%-12/16 
%{\color{blue}
%(Sorry but I don't understand the following sentence.
%Can we break it into a few sentences?)
%}
%YM-12/16
We will see that the expansion around the outgoing Vaidya metric agrees with 
the result from another perturbative calculation for $r - a = \mathcal O(\kappa)$. 
We also find that the integration constants are related to the validity of the expansion 
at higher orders and should be chosen carefully. 
%-12/16
%
%We will see that the leading order of the perturbation for $r - a = \mathcal O(\kappa)$ 
%is reproduced by the perturbation around the outgoing Vaidya metric with first-order corrections 
%which satisfy the consistency condition of the perturbation to second order. 
%%%% PM +
%(The subtlety behind this calculation will be clarified in Sec.\ref{sec:vaidya-neck}.)
% YM: Another expansion above means the calculation in Sec.4.3 
% YM: so it sounds strange that its subtlety is discussed in the same section, Sec.4.3. 
%%%% PM -
%through a careful analysis,
%it is actually possible to extract reliable information
%about the geometry around the neck.
The advantage of this approach is that
its description about the geometry around the neck
is smoothly continued to the asymptotic region at large distance.

\begin{figure}
\begin{center}
 \includegraphics[scale=0.4]{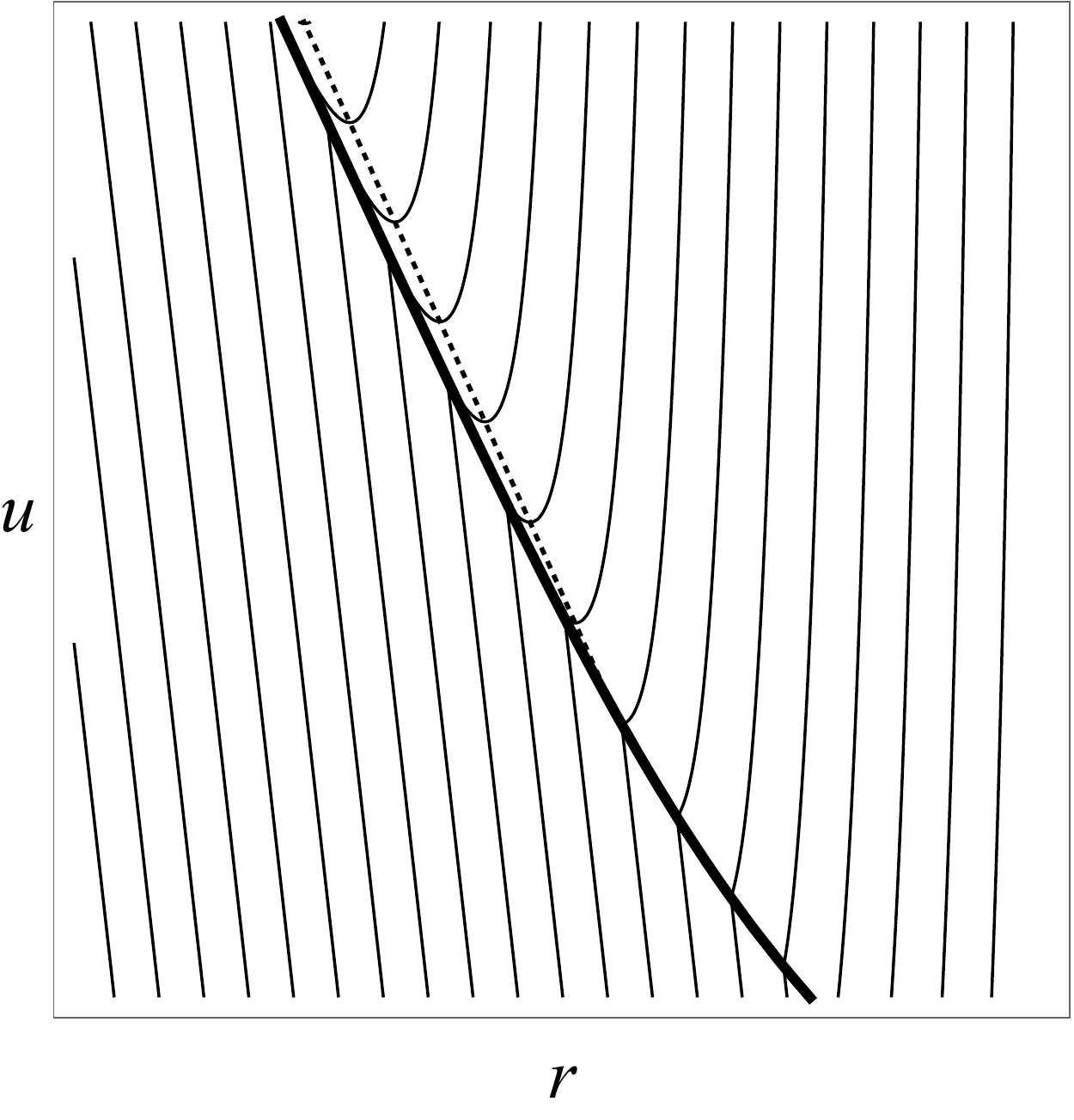}
 \caption{\small
 Lines of $R(u,r)=\text{const.}$ in the $(u,r)$-coordinates. 
 The bold line is the collapsing thin shell, and the dotted line is the neck, 
 or equivalently, the apparent horizon. 
 The distance between the collapsing shell and apparent horizon 
 is very small in these coordinates. 
 Strictly speaking, the distance in these coodinates is different from the proper distance, 
 it can be a very rough estimate since $f(u,r)$ in \eqref{vaidya} is small near the apparent horizon. 
 More precise arguments are given in Sec.~\ref{sec:distance-pert} and in Sec.~\ref{sec:distance-deep}
 }\label{fig:contour_v}
\end{center}
\end{figure}

The main results of this section to be derived below are the following. 
%YM-4/8
The solution of the exterior geometry in the form of \eqref{vaidya} is obtained as 
%-4/8
\begin{align}
 f(u,r) 
 &\simeq 
 1 - \frac{a(u)}{r} + \cdots \ , 
 \label{sol-f}
 \\
 R(u,r) 
 &\simeq 
 r - \frac{\kappa}{4 a(u)} \log\left(\frac{r - a(u) + 2 a(u) \dot a(u)}{a(u)}\right) + \cdots \ , 
 \label{sol-R}
\end{align}
where higher-order corrections indicated by ``$\cdots$'' are of $\mathcal O(\kappa)$ 
%YM-11/12
for $r - a = \mathcal O(a)$ (far outside the apparent horizon) 
%-11/12
but $\mathcal O(\kappa^2)$ for $r -a = \mathcal O(\kappa)$ (around the apparent horizon). 
%YM-4/8
The geometry is given by connecting the solution above with the flat spacetime 
at the collapsing shell (See Fig.~\ref{fig:contour_v}. 
%-4/8
The areal radius $R$ increases as it goes further inside the apparent horizon $r = a(u)$
%%%PM
%YM-12/16
(as the value of $v$ gets smaller on a constant-$u$ curve), 
while it decreases along the collapsing shell
(as the value of $u$ gets larger on a constant-$v$ curve).
%-12/16
%%%MP 
%The $r$-coordinate of the neck is
%(see eq.\eqref{neck-2})
%\begin{equation}
% r = a(u) - 2 a(u) \dot a(u) + \frac{\kappa}{4 a(u)} \ ,
%\end{equation}
%and the $r$-coordinate of the collapsing shell approaches the value 
%\begin{equation}
% r = a(u) - 2 a(u) \dot a(u) \ .
%\end{equation}
%While the difference in their $r$-coordinate is $\mathcal{O}(\kappa)$,
The rate of decrease of the areal radius $R_s$ along the collapsing shell 
is much smaller than that of the Schwarzschild radius. 
We will see below that,
although the difference in their areal radii $R$ will be as large 
as $\mathcal{O}(a)$, as the shell goes to deeper region inside the apparent horizon, 
the shell is only separated from the neck 
by a proper distance of the order of a Planck length.

%Although we study the difference of the areal radius 
%and the proper distance, between the collapsing shell and apparent horizon, 
%by using the perturbation around the Vaidya metric, 
%the expressions
While eqs.\eqref{sol-f} and \eqref{sol-R} are not very useful
deep inside the neck where $r - a = \mathcal O(a)$,
we will study the collapsing shell deep inside the neck 
in more detail in Sec.\ref{sec:deep-dyn}. 

%We will extend this result in the next section
%to apply to collapsing shells already deep inside the neck.

%-10/10

\subsection{Perturbation of Vaidya Metric}

We first take into account the effect of the Hawking radiation by introducing 
the outgoing energy at the leading order of the expansion,
so that the solution of the semi-classical Einstein equation at the leading order 
is given by the outgoing Vaidya metric. 
Then we treat the remaining part of the vacuum energy-momentum tensor
(including the incoming negative vacuum energy flux) as a perturbation.

The energy-momentum tensor \eqref{Tuu-vac}--\eqref{Tthth-vac}
is fixed by the trace anomaly and the conservation law.
For the metric \eqref{vaidya},
it is equivalent to
\begin{align}
 \left\langle T_{uu}\right\rangle 
 &= 
 \frac{1}{R^2} \left[\beta + \gamma + f \partial_r^2 f - \frac{1}{4} \left( \partial_r f\right)^2\right] 
 + \mathcal O(\kappa) \ , 
 \label{Tuu0}
\\
 \left\langle T_{rr}\right\rangle 
 &= 
 \frac{1}{R^2 f^2} \left[2 \gamma + f \partial_r^2 f - \frac{1}{2} \left( \partial_r f\right)^2\right]
 + \mathcal O(\kappa) \ , 
 \label{Trr0}
\\
 \left\langle T_{ur}\right\rangle 
 &= 
 \left\langle T_{ru}\right\rangle 
 = 
 \frac{1}{R^2 f} \left[ \gamma + f \partial_r^2 f - \frac{1}{4} \left( \partial_r f\right)^2\right] 
 + \mathcal O(\kappa) 
 \ . 
 \label{Tur0}
\end{align}

We solve the semi-classical Einstein equation
order by order in the $\kappa$-expansion. 
We expand $f(u,r)$ and $R(u,r)$ as 
\begin{align}
 f(u,r) 
 &= 
 f_0(u,r) + \kappa f_1(u,r)
 + \kappa^2 f_2(u, r) + \mathcal O(\kappa^3) \ , 
\\
 R(u,r) 
 &= 
 R_0(u,r) + \kappa R_1(u,r)
 + \kappa^2 R_2(u, r) + \mathcal O(\kappa^3) \ . 
\end{align}
At the leading order, the semi-classical Einstein equation is solved by
\begin{align}
 R_0(u,r) &= R_{00}(u) + R_{01}(u) r \ , 
\\
 f_0(u,r) &= R_{01}^{-2}(u) - \frac{a(u)}{R_0(u,r)} \ ,  
\end{align}
where $R_{00}(u)$, $R_{01}(u)$ and $a(u)$ are integration constants. 

In the lowest-order approximation, 
$a(u)$ corresponds to the Schwarzschild radius. 
The other integration constants $R_{00}(u)$ and $R_{01}(u)$ 
can be absorbed by the redefinition of the $(u,r)$-coordinates. 
Note that $u$-dependence is very slow and contribute only to 
the higher order corrections, for example, $\dot a(u) = \mathcal O(\kappa)$. 
We choose the gauge condition $R_{00}(u) = 0$ and $R_{01}(u) = 1$
to obtain the standard expression of the outgoing Vaidya metric; 
\begin{align}
 R_0(u,r) &= r \ , 
 \label{R0vaidya}
\\
 f_0(u,r) &= 1 - \frac{a(u)}{r} \ . 
 \label{f0vaidya}
\end{align}

At the next-to-leading order, the semi-classical Einstein equation is solved by
\begin{align}
 R_1(u,r) 
 &= 
 R_{10}(u) + R_{11}(u) r 
\notag\\&\quad
 + \frac{1}{8 r a^2(u)} 
 \biggl[
  a(u)^2 \left(8 \gamma(u)  r^2-3\right) -2 r(r-2 a(u)) \log (r) 
\notag\\&\qquad\qquad\qquad\quad
 -2 r (r-2 a(u)) \left(4 \gamma(u)  a(u)^2-1\right) \log (r-a(u)) 
 \biggr] \ , 
 \label{R1vaidya}
\\
 f_1(u,r) 
 &= 
 \frac{a_1(u)}{r} 
\notag\\&\quad
 + 
 \frac{1}{2 r^3 a^2(u)} 
 \biggl[a^2(u) \left(r -a(u) -2 r^3 \gamma (u)\right) 
\notag\\&\qquad\qquad\qquad\quad
 + 2 r a(u)^2 \left(a(u) R_{10}(u) -2 r^2 R_{11}(u)\right) 
\notag\\&\qquad\qquad\qquad\quad
  + \left(-2 r^2 a(u)+r a(u)^2+r^3\right) \log r 
\notag\\&\qquad\qquad\qquad\quad
  + r (r-a(u))^2 \left(4 a(u)^2 \gamma (u)-1\right) \log (r-a(u))
 \biggr] \ , 
\label{f1vaidya}
\\
 \dot a(u) 
 &= - \kappa \beta(u) \ , 
\label{HawkingVaidya}
\end{align}
where $R_{10}(u)$, $R_{11}(u)$ and $a_1(u)$ are the integration constants,
%%% PM +
which will be set to $0$'s as they
can be absorbed by the parameters
$R_{00}(u)$, $R_{01}(u)$ and $a(u)$
at the leading order, respectively. 
%%% PM -

The first-order solution behaves in the asymptotic region ($r\to\infty$) as 
\begin{align}
 R_1(u,r) 
 &= 
 \gamma \left(r-r\log r\right) + \mathcal O(r^0) \ , 
 \\
 f_1(u,r) 
 &= 
 - \gamma \left(1-2\log r\right) + \mathcal O(r^{-1}) \ . 
 \label{f1-limit}
\end{align}
By imposing $\gamma=0$ \eqref{gamma=0}, which corresponds 
to the condition of no incoming energy in the past null infinity, 
eqs.\eqref{R1vaidya} and \eqref{f1vaidya} are smaller than the leading-order solutions 
eqs.\eqref{R0vaidya} and \eqref{f0vaidya}, respectively, 
and hence the perturbative expansion is valid for large $r$. 

On the other hand,
the first-order corrections \eqref{R1vaidya} and \eqref{f1vaidya} 
behave near the Schwarzschild radius $r = a(u)$ as 
\begin{align}
 R_1(u,r) 
 &= 
 - \frac{1}{8 a(u)} \left[3 + 2\log\left(\frac{r-a(u)}{a(u)}\right)\right] + \mathcal O(r-a(u)) \ , 
 \\
 f_1(u,r) 
 &= 
 \frac{r-a(u)}{2a(u)^3} + \mathcal O\left(\left(r-a(u)\right)^2\right) \ . 
\end{align}
For $f(u,r)$, the first-order correction is smaller than 
the leading-order solution and the expansion is valid even around $r=a$. 
The negative vacuum energy introduces a small correction:
\begin{equation}
 f(u,r) = f_0(u, r) + f_1(u, r) + \cdots
 = \left(\frac{1}{a(u)} + \frac{\kappa}{2 a^3(u)}\right)\left(r-a(u)\right) 
 + \mathcal O((r-a(u))^2) \ . 
\end{equation}
However, the correction term $R_1(u,r)$ for the radius $R(u,r)$ 
has a logarithmic divergence at $r=a(u)$. 

The areal radius $R(u,r)$ has a local minimum slightly outside the point $r=a(u)$
where $R_1(u,r)$ approaches positive infinity.
The local minimum of $R$ (the neck) is located at the point where
\begin{equation}
 0 = \partial_r R(u,r) \simeq 1 - \frac{\kappa}{4a(u)[r-a(u)]} + \cdots \ , 
\end{equation}
which is solved by
\begin{equation}
 r = a(u) + \frac{\kappa}{4 a(u)} + \mathcal O(\kappa^2) \ .
 \label{neck-VP}
\end{equation}
This is consistent with the structure we have seen in previous sections.

Now, we consider a collapsing shell at the speed of light. 
The shell is located on an incoming null line $r = r_s(u)$, 
which is a solution of the differential equation 
\begin{equation}
 \frac{dr_s}{du} = - \frac{1}{2} f(u,r_s) \ . 
\end{equation}
%Although the linear order correction of $\kappa$-expansion in the radius $R(u,r)$ 
%becomes significant as it approaches to $r\sim a(u)$, 
%it is irrelevant to the position of the shell in $(u,r)$-coordinates. 
%As the correction for $f(u,r)$ is still small near $r=a(u)$, 
%we can just ignore them. 
When the shell approaches the Schwarzschild radius
($r_s - a(u) = \mathcal O(\kappa)$), 
the position of the shell $r_s(u)$ is approximately given by
\begin{equation}
 r_s(u) = a(u) + s e^{-\frac{u}{2a(u)}} - 2 a(u) \dot a (u) + \mathcal O(\kappa^2) \ , 
 \label{shell-vaidya}
\end{equation}
where $s$ is an integration constant. 
Recall that the time evolution of $a(u)$ is slow; $\dot a(u) = \mathcal O(\kappa)$. 
As the second term is exponentially suppressed over time, 
the shell approaches 
%YM-10/10
\begin{equation}
 r \simeq a(u) - 2 a(u) \dot a (u) + \mathcal O(\kappa^2) \ . 
 \label{rs-limit}
\end{equation}
This is slightly outside the point $r=a(u)$ as 
$a(u)$ is decreasing. 
The shell never reaches the value in eq.\eqref{rs-limit} within finite time, 
unless $s=0$ in eq.\eqref{shell-vaidya}. 
The null surface \eqref{rs-limit} can be interpreted as the past horizon 
of the exterior geometry.
\footnote{%
The black-hole geometry for a collapsing null shell is obtained by 
connecting this exterior geometry with flat spacetime on the collapsing shell with $s>0$.
%%% PM +
The past horizon \eqref{rs-limit} does not really exist
since this region of the exterior geometry is replaced by the flat spacetime.
%%% PM -
} 
%-10/10

%The time evolution of the Schwarzschild radius $a(u)$ 
%is determined by \eqref{HawkingVaidya}. 
The Hawking radiation $\beta(u)$ can be calculated 
from the junction condition 
%of the energy-momentum tensor 
on the collapsing shell. 
%As we have discussed in Sec.~\ref{sec:neck-shell}, 
%the outgoing energy must be continuous at the null shell. 
%Since it is zero in the flat spacetime inside the shell, 
%it should be zero too just outside of the shell. 
The condition is given by $\langle T_{uu}\rangle = 0$ in $(u,v)$-coordinates, 
that is,
\begin{align}
 0 &= \langle T_{uu}(u,v)\rangle 
 \nonumber
\\
 &= \langle T_{uu}(u,r_s)\rangle 
 + 2 \left(\frac{\partial r}{\partial u}\right) \langle T_{ur}(u,r_s)\rangle 
 + \left(\frac{\partial r}{\partial u}\right)^2 \langle T_{rr}(u,r_s)\rangle 
 \nonumber
\\
 &= \beta(u) + \frac{a (3a-4r_s)}{8 r_s^4} + \mathcal O(\kappa) \ . 
\end{align}
As eq.\eqref{rs-limit} implies that $r_s-a(u) = \mathcal O(\kappa)$, 
the Hawking radiation is estimated as 
\begin{equation}
 \beta(u) = \frac{1}{8 a^2(u)} + \mathcal O(\kappa) \ ,
\end{equation}
in agreement with eq.\eqref{Hawking-neck}.
Thus, the time derivative of the Schwarzschild radius $\dot a(u)$ 
is given by \eqref{HawkingVaidya}
\begin{equation}
 \dot a (u) = - \kappa \beta(u) = - \frac{\kappa}{8a^2(u)} \ .
\end{equation}
According to eq.\eqref{rs-limit},
the position of the shell approaches
\begin{equation}
 r_s = a(u) + \frac{\kappa}{4 a(u)} \ . 
\end{equation}
%YM-10/10
%Interestingly,
This is nothing but the $r$-coordinate \eqref{neck-VP}
of the neck.
%i.e. the local minimum of the areal radius $R(u,r)$.

Naively, this result seems to imply that 
%%% PM +
the $r$-coordinate at the collapsing shell
differs from that at the apparent horizon
only by $\mathcal{O}(\kappa^2)$
after it gets inside the neck. 
%%% PM -
However, this naive expectation is incorrect
for our perturbative expansion.
%YM-10/21
%%% PM +
%The integration constant $R_{10}(u)$ 
%is related to the small correction to the coordinate $r$. 
%In this section, we have chosen $R_{10}(u) = 0$ by an appropriate redefinition of $r$, 
%another choice of $R_{10}(u)$ changes the position of $f(u,r) = 0$ in the $r$-coordinate. 
%Of course, this is not physical, but simply because of the change of the definition of $r$. 
%[The following is removed because it is essentially the same
%as the paragraph in the beginning of the next subseciton.]
%However, the change of the position of the logarithmic divergence in $R(u,r)$ 
%cannot be read off from \eqref{R1vaidya} 
%because it appears first in $R_1(u,r)$ and its correction will be at the second order. 
%This also implies that the position of the logarithmic divergence of $R(u,r)$ 
%possibly has non-trivial correction at second order. 
%Although the position of $f(u,r)=0$ and that of 
%logarithmic divergence in $R(u,r)$ are seemingly at the same position, $r=a$, 
%the relative position might have difference of $\mathcal O(\kappa)$. 
%The position of $f(u,r)=0$ and that of the logarithmic divergence of $R(u,r)$ 
%are related to that of the past horizon and that of the neck, respectively. 
To find more precisely the relative position between the collapsing shell and the apparent horizon, 
we have to include the second-order effect. 
%-10/21
%It should be noted that the relation is not exact as
%both the position of the shell and that of the local minimum of $R(u,r)$ 
%have higher-order corrections. 
%so of course the shell does not always stay exactly on top of the neck.
%But this result does show that,
%surprisingly,
%the surface of the shell is always very close to the neck
%in terms of the $r$-coordinate,
%regardless of how long the shell has collapsed.
In the next subsection,
we will study the second-order correction
%to estimate the order of magnitude of this small separation.
to argue that the difference between the areal radius at the shell and 
that at the apparent horizon at the same $u$ increases with $u$.
%-10/10
%%% PM -

%%%%%%%%%%%%%%%%%%%%%%%%%%%%%%%%%%%%%%%%%%%%%%%%%%%%%%%%%%%%%%%%%%%%%%%%%%%%%%%%
%%%%%%%%%%%%%%%%%%%%%%%%%%%%%%%%%%%%%%%%%%%%%%%%%%%%%%%%%%%%%%%%%%%%%%%%%%%%%%%%
%%%%%%%%%%%%%%%%%%%%%%%%%%%%%%%%%%%%%%%%%%%%%%%%%%%%%%%%%%%%%%%%%%%%%%%%%%%%%%%%

\subsection{Higher-Order Correction}
\label{sec:vaidya-2nd}

In the perturbative expansion around the Vaidya metric \eqref{vaidya}, 
the first-order correction of the areal radius $R$ is
%of the same order of magnitude as 
not much smaller than the leading order term
in the region where $r - a \sim \mathcal O(\kappa)$. 
%YM-10/10
In addition,
another issue with the first-order approximation is the following.
The divergence of the first-order correction to the areal radius $R$ at $r=a$
%%% PM +
looks harmless as it is located inside the (non-existent) past horizon,
hence outside the region of interest.
This is because the past horizon
%%% PM -
is moved by the first-order correction slightly outward
to $r = a - 2 a \dot a > a$. 
%YM-10/21
%YM-11/12
However, 
this difference is only of $\mathcal O(\kappa)$,
and the position of the divergence in $R$ can be read off from eq.\eqref{R1vaidya}
only up to an $\mathcal O(\kappa)$-correction,
so the statement above about the divergence is not reliable.
To confirm the relative position of the divergence in $R$ with respect to the past horizon, 
the second-order correction must be studied. 
%-11/12

In this subsection, 
we investigate the second-order correction.
The second-order correction to $R$ turns out to have a pole at $r=a$, 
which is comparable to the first order term, and 
naively implies the breakdown of the perturbative expansion. 
However, we can absorb the pole in the second-order term 
by a shift of the logarithmic divergence in the 1st-order term. 
This implies that the pole at the second order can be canceled  
by using an appropriate redefinition of the coordinate $r$, 
or equivalently, by choosing integration constants at the first order appropriately. 
%By using the appropriate choice of the integration constants at first order, 
%the second order terms have no divergence and then sufficiently smaller than the first order terms. 
%%% PM +

%YM-11/12
This is in fact related to the relative position between 
the past horizon \eqref{rs-limit} and the logarithmic divergence of $R$ in eq.\eqref{R1vaidya}. 
The relative position at $\mathcal O(\kappa)$ is related to 
the validity of the perturbative expansion at the second order. 
As the leading-order term of the difference between the $r$-coordinates of 
the collapsing shell and the apparent horizon is $\mathcal O(\kappa)$, 
the second-order correction is important to remove the ambiguity
of the position at $\mathcal O(\kappa)$. 
The third-order correction should be taken into consideration to 
calculate the $\mathcal O(\kappa^2)$-correction to the positions, etc.
To justify the assumption that the divergence in higher-order terms
can be canceled by using an appropriate definition of the $r$-coordinate, 
we will consider another perturbative expansion
%In this subsection, 
%we investigate the second-order correction.
%%and show that the divergence of $R$ is located on the past horizon. 
%%-10/21
%Although the 2nd-order terms in $R$ turn out to be
%larger than the 1st-order terms, 
%we can absorb the dominant 2nd-order term
%by a shift of the logarithmic divergence in the 1st-order terms. 
%%Such a manipulation may seem unreliable,
%%as the 3rd-order terms may also be larger than the 2nd-order terms.
%We will show that this worry is unnecessary
%via a calculation in a different coordinate system
in Sec.\ref{sec:vaidya-neck}.
In the end,
the second-order result in this subsection
allows us to smoothly connect the solution of the metric
in the asymptotically flat region
to the near-horizon region.
%%% PM -
%-10/10
%-11/12

%We first consider the first-order correction for the energy-momentum tensor. 
The expectation value of the energy-momentum tensor is expanded as 
\begin{equation}
 \langle T_{\mu\nu}\rangle 
 = T^{(0)}_{\mu\nu} + \kappa T^{(1)}_{\mu\nu} 
 + \mathcal O(\kappa^2) \ . 
\end{equation}
The leading-order terms $T_{\mu\nu}^{(0)}$ 
are given by \eqref{Tuu0}-\eqref{Tur0}, 
in which the $u$-derivatives are ignored as higher-order terms. 
The linear-order corrections $T^{(1)}_{\mu\nu}$ contains 
the effects of $u$-derivatives of leading-order terms. 
By using the solution at the leading order \eqref{R0vaidya}-\eqref{f0vaidya},
$T^{(1)}_{\mu\nu}$ is calculated via the conservation law
at the first order of the $\kappa$-expansion as 
\begin{align}
 T^{(1)}_{uu}
 &=
 \beta_1(u) + \frac{a(u) \dot a(u)}{2 \kappa r^2 \left(r-a(u)\right)} 
\notag\\&\quad
 - \frac{2a(u)}{r^3} f_1(u,r) - \frac{a(u)}{2r^2} \partial_r f_1(u,r) 
 \left(1- \frac{a(u)}{r}\right) \partial_r^2 f_1(u,r) 
\ , 
 \\
 T^{(1)}_{rr}
 &=
 \frac{r}{r-a(u)} \gamma_1(u)
 + \frac{2 r \dot a(u)}{\kappa \left(r-a(u)\right)^3} 
\notag\\&\quad
 + \frac{a(u) (2 r-a(u))}{r(r-a(u))^3} f_1(u,r) 
 - \frac{a(u)}{(r-a(u))^2} \partial_r f_1(u,r) 
 + \frac{r}{r-a(u)} \partial_r^2 f_1(u,r) 
 \ , 
 \\
 T^{(1)}_{ur} 
 &= 
 T^{(1)}_{ru} 
 = 
 \frac{2r^2}{\left(r-a(u)\right)^2} \gamma_1(u) 
 + \frac{\dot a(u)}{\kappa \left(r-a(u)\right)^2} 
\notag\\&\quad
 + \frac{a^2(u)}{4r^2(r-a(u))^2} f_1(u,r) 
 - \frac{a(u)}{2r(r-a(u))^2} \partial_r f_1(u,r) 
 + \partial_r^2 f_1(u,r) 
 \ ,  
\end{align}
where $\beta_1$ and $\gamma_1$ are integration constants. 
The integration constant $\beta_1$ corresponds to 
the linear-order correction to the Hawking radiation. 
The other constant $\gamma_1$ is the linear-order correction to 
the incoming energy at the past infinity;
we should choose $\gamma_1=0$. 

Solving the semi-classical Einstein equation at the second order,
there is a divergence in the second-order term of $R$.
It is
\begin{equation}
 R_2(u,r) = - \frac{\dot a(u)}{2\kappa(r-a(u))} + \mathcal O((r-a(u))^0) \ , 
 \label{R2div}
\end{equation}
%YM-10/21
which is large compared with the first-order correction %gives an $\mathcal O(\kappa)$-contribution 
%-10/21
to the areal radius $R$
if $r - a(u) = \mathcal O(\kappa)$. 
Together with the logarithmic term in $R_1(u,r)$, 
\begin{equation}
 R_1(u,r) = - \frac{1}{4a(u)} \log (r-a(u)) + \cdots \ , 
\end{equation}
the pole in $R_2(u,r)$ amounts to a shift of the location of
the logarithmic divergence in $R_1(u,r)$ as
\begin{equation}
 - \frac{1}{4a(u)} \log (r-a(u)) - \frac{\dot a(u)}{2(r-a(u))} 
 \simeq - \frac{1}{4a(u)} \log \left(r-a(u) + 2 a(u) \dot a(u)\right)) \ . 
 \label{log-corr}
\end{equation}

%YM-10/21
Although the pole in eq.\eqref{R2div} can be interpreted as 
a small shift of the logarithmic divergence in $R_1(u,r)$, 
it seemingly breaks down the perturbative expansion since 
it becomes too large as a second-order correction. 
This interpretation above of the 2nd-order term \eqref{R2div}
can be justified more precisely as follows. 
To avoid the divergence in $R_2(u,r)$, 
we need to keep the integration constant $R_{10}(u)$ in eq.\eqref{R1vaidya}
as an arbitrary function of $u$,
and then $R_2(u,r)$ would be found to be 
\begin{equation}
 R_2(u,\tilde r) 
 = - \frac{R_{10}(u) + 2 a(u) \dot a(u)}{4 a(u) \kappa(\tilde r-a(u))} + \mathcal O((\tilde r-a(u))^0) \ ,  
 \label{R2div2}
\end{equation} 
where we denoted the $r$-coordinate as $\tilde r$,
in order to distinguish from the $r$-coordinate used in the $R_{10}(u) = 0$ gauge. 
Different choices of $R_{10}(u)$ corresponds to different definitions of the $r$-coordinates 
which are related to each other by coordinate transformations, $r \to \tilde r = r - R_{10}(u)$. 
The integration constant $R_{10}(u)$ should be chosen such that 
the pole in \eqref{R2div2} cancels; 
\begin{equation}
 R_{10}(u) = - 2 a(u) \dot a(u) \ . 
 \label{R10}
\end{equation}
Then, the second-order correction is negligible and the perturbation is valid. 
By using \eqref{R10}, or equivalently $\tilde r$, 
the position of the logarithmic divergence is not shifted away from $\tilde r = a(u)$. 
%YM-10/10
Since the $r$-coordinate in the $R_{10}(u)=0$ gauge is related to $\tilde r$,
the $r$-coordinate in the gauge \eqref{R10}, 
via the relation $r = \tilde r + R_{10}(u)$,
the position of the divergence is indeed shifted to
\begin{equation}
 r = a(u) - 2 a(u) \dot a(u) \ . 
\end{equation}

On the other hand,
$f(u,r)$ does not have 
the divergence in the second-order correction. 
By using eq.\eqref{R10}, instead of $R_{10}(u)=0$, 
eq.\eqref{f1vaidya} gives $\mathcal O(\kappa)$ correction to $f(u,\tilde r)$ as 
\begin{equation}
 f(u,\tilde r) = \tilde r - a(u) - 2 a(u) \dot a(u) + \mathcal O(\kappa^2) \ , 
\end{equation}
for $\tilde r - a(u) = \mathcal O(\kappa)$, 
which is equivalent to $f(u,r) = r - a(u) + \mathcal O(\kappa^2)$ after 
the coordinate transformation to the $R_{10}(u)=0$ guage.
%$r = \tilde r + R_{10}(u)$. 

Similarly, the position of the past horizon of the exterior geometry 
is placed at the same position \eqref{rs-limit},
up to $\mathcal{O}(\kappa^2)$.
%-10/21
The local minimum of the radius $R$ is now placed at 
\begin{equation}
 r = a(u) - 2 a(u) \dot a(u) + \frac{\kappa}{4 a(u)} \ ,
 \label{neck-2}
\end{equation}
as a second-order correction to eq.\eqref{neck-VP}.

%YM-10/21
In this subsection, we have seen that 
the $r$-coordinate should be chosen appropriately for the validity of the perturbation. 
The solutions \eqref{sol-f} and \eqref{sol-R} can be calculated 
at the first order with an appropriate choice of the integration constant, 
while the appropriate choice is found by studying the second-order terms. 
%the second order corrections 
%to the areal radius $R$ are important to see the relative position  
%near the apparent horizon, $r - a = \mathcal O(\kappa)$. 
%On the other hand, all correction terms for $f(u,r)$ is smaller than the leading order. 
Since the second-order corrections are important to see 
the relative position of the past horizon and the logarithmic divergence of the areal radius
to $\mathcal O(\kappa)$ corrections,
we expect that the third-order corrections would be necessary only for 
$\mathcal O(\kappa^2)$ correction to the relative position. 
However, 
as there can be higher-order divergences in higher-order terms,
it is not totally obvious that all higher-order corrections
will never introduce a large correction to the first-order solution at  $r - a = \mathcal O(\kappa)$ 
in the perturbative expansion around the Vaidya metric. %, 
%which is justified only for $r - a = \mathcal O(a)$. 

%%% PM +
In the next subsection,
we consider another expansion focused on the neighborhood of $r - a = \mathcal O(\kappa)$, 
and justify our claim that the corrections to the second order are sufficient to 
describe the geometry near and outside the apparent horizon,
so that the Vaidya-like metric \eqref{vaidya} allows us to describe the geometry
in the near-horizon region $r - a = \mathcal{O}(\kappa)$ 
as well as in the asymptotically flat region
with good approximation.

\subsection{Perturbation Near The Neck}
\label{sec:vaidya-neck}

%In the previous section, we have seen that 
%the areal radius $R$ has a local minimum 
%due to the back-reaction of the vacuum energy.
In the perturbative expansion around the Vaidya metric
discussed above, 
it is implicitly assumed that $r - a(u) \sim \mathcal O(\kappa^0)$, 
and it is not totally clear if the expansion is valid for
the smaller neighborhood where
$r-a(u) \sim \mathcal O(\kappa)$,
as we have mentioned above.
To clarify the subtlety involved,
we consider another perturbative expansion 
which is good around the local minimum of $R$, 
and show that the results in the previous subsection is 
indeed valid in this region. 

Instead of eq.\eqref{vaidya},
we consider the following metric:
\begin{equation}
 ds^2 = - \tilde f(u,z) du^2 - 2 \kappa du\,dz + R^2(u,z) d \Omega^2 \ . 
 \label{vaidya-neck}
\end{equation}
Here, we focus on the small neighborhood around the neck. 
%The radial coordinate is chosen such that 
%\begin{equation}
% dz \sim \kappa dr \ , 
%\end{equation}
%and hence, $(u,z)$-component of the metric is of $\mathcal O(\kappa)$. 
We identify the radial coordinate $z$
with the $r$-coordinate in eq.\eqref{vaidya} via
%by using $r$ in \eqref{vaidya} as 
\begin{equation}
 r = a(u) + \kappa z \ , 
 \label{z-vaidya}
\end{equation}
and so $\tilde f(u,z)$ is related to $f(u,r)$ in eq.\eqref{vaidya} as 
\begin{equation}
 \tilde f(u,z) = f(u,r) + 2 \dot a(u) \ . 
 \label{tilde-f}
\end{equation}
Around $r=a(u)$, we have
%$\tilde f(u,z)$ becomes small as 
\begin{equation}
 \tilde f(u,z) = \mathcal O(\kappa) \ . 
\end{equation}
At the neck,
\begin{equation}
 R(u,z) = a(u) + \mathcal O(\kappa) \ . 
\end{equation}

We consider the following expansion:
\begin{align}
 \tilde f(u,z) 
 &= \kappa \tilde f_0(u,z) + \mathcal O(\kappa^2) \ , 
 \\
 R(u,z) 
 &= 
 a(u) + \kappa \tilde{R}_0(u,z) + \mathcal O(\kappa^2) \ . 
\end{align}
The leading-order terms of the semi-classical Einstein equation give
the following differential equations for $\tilde f_0(u,z)$ and $\tilde{R}_0(u,z)$; 
\begin{align}
 0 
 &= 
 \left(\partial_z \tilde f_0(u,z)\right)^2 
 - 2 \tilde f_0(u,z) 
  \left(
   \partial_z^2 \tilde f_0(u,z) + 2 a(u) \tilde f_0(u,z) \partial_z^2 \tilde{R}_0(u,z) 
  \right)
 \ , 
 \\
 0 
 &= 
 \left(\partial_z \tilde f_0(u,z)\right)^2 
 - 4 \tilde f_0(u,z) 
  \left[
   1 - a(u) \left(\partial_z \tilde f_0(u,z)\right) \left( \partial_z \tilde{R}_0(u,z)\right)  
  \right] 
 \ , 
 \\
 0 
 &= 
 \kappa \beta(u) + a(u) \dot a(u) \partial_z \tilde f_0(u,z) \ , 
 \label{HawkingVN}
\end{align}
where we have again assumed that
the $u$-derivative is small: $\partial_u \sim \mathcal O(\kappa)$. 

One of the solutions of the differential equations above is trivial |  
the first-order correction of $R$ is constant in the radial direction,
$\tilde{R}_0(u,z) = \tilde{R}_0(u)$. 
The other non-trival solution is 
\begin{align}
 \tilde f_0(u,z) 
 &= 
 \tilde{f}_{00}(u) + \tilde{f}_{01}(u) z 
 \ , 
 \label{f-vn}
 \\
 \tilde{R}_0(u,z) 
 &= 
 \tilde{R}_{00}(u) + \frac{z}{a(u)f_{01}(u)} - \frac{1}{4a(u)} \log \tilde f_0(u,z) \ , 
 \label{r-vn}
\end{align}
where $\tilde{f}_{00}(u)$, $\tilde{f}_{01}(u)$ and $\tilde{R}_{00}(u)$
are arbitrary functions. 
This result is consistent with the expansion around the Vaidya metric, 
%YM-10/10
%\eqref{f1vaidya} and \eqref{R1vaidya} for large $z$.
eqs.\eqref{sol-f}--\eqref{sol-R}. 
By using eqs.\eqref{z-vaidya} and \eqref{tilde-f}, 
eqs.\eqref{sol-f} and \eqref{sol-R} give the same expressions
as eqs.\eqref{f-vn} and \eqref{r-vn}. 
The integration constants are fixed by the relation 
to eqs.\eqref{sol-f} and \eqref{sol-R} as 
%-10/10
\begin{align}
 \tilde f_0(u,z) 
 &= 
 2 \kappa^{-1}\dot a(u) + \frac{z}{a(u)} \ ,  
 \\
 \tilde{R}_0(u,z) 
 &= 
 - \frac{3}{8a(u)} + z 
 - \frac{1}{4a(u)} \log \left( 2\kappa^{-1} \dot a(u) + \frac{z}{a(u)}\right) \ . 
\end{align}
Eq.\eqref{HawkingVN} implies that 
the Hawking radiation satisfies
\begin{equation}
 \dot a(u) = - \kappa \beta(u)  \ . 
\end{equation}
%YM-10/10
%The neck is located at
%\begin{equation}
% z = - 2 \kappa^{-1} a(u) \dot a(u) + \frac{1}{4a(u)} \ ,
%\end{equation}
%where $\partial_z \tilde{R}_0(u,z) = 0$.
%In terms of the $r$-coordinate, 
%the neck is at
%\begin{equation}
% r = a(u) - 2 a(u) \dot a(u) + \frac{\kappa}{4 a(u)} \ . \label{neck-VN}
%\end{equation}
%This is exactly the same as the result \eqref{neck-2}
%in the previous section.
%
%The position of the shell $z_s$ is given by the solution of the differential equation, 
%\begin{equation}
% \frac{dz_s}{du} = - \frac{1}{2} \tilde f(u,z_s) \ , 
%\end{equation}
%which is solved as 
%\begin{equation}
% z_s = - 2 \kappa^{-1} a(u) \dot a(u) + C e^{-\frac{u}{2a(u)}} \ . 
%\end{equation}
%Thus, the collapsing shell approaches $z = -2 \kappa^{-1} a \dot a$,
%or equivalently, 
%\begin{equation}
% r = a(u) - 2 a(u) \dot a(u) \ , 
% \label{nh-ph}
%\end{equation}
%which is the past horizon of the exterior geometry. 
%The areal radius  diverges at \eqref{nh-ph}, 
%while the areal radius at the collapsing shell is still decreasing, 
%as we have seen in the previous subsection. 
Therefore,
we have verified that the expressions \eqref{sol-f} -- \eqref{sol-R} 
are indeed reliable even near the apparent horizon $r - a = \mathcal O(\kappa)$. 

Now we consider the time-evolution of the collapsing shell using eqs.\eqref{sol-f} and \eqref{sol-R}. 
%Although the shell asymptotically approaches \eqref{rs-limit}, 
%the areal radius at the shell does not increase but decreases. 
Since the higher-order corrections to $f(u,r)$ is negligible even for $r - a = \mathcal O(\kappa)$, 
the locus of the collapsing shell is given by eq.\eqref{shell-vaidya}, 
and the areal radius $R$ there is calculated as 
\begin{align}
 R_s(u) 
 &\simeq
 r_s(u) - \frac{\kappa}{4 a(u)} \log\left(\frac{r_s(u) - a(u) + 2 a(u) \dot a(u)}{a(u)}\right) 
 + \cdots 
 \\
 &\simeq
 a(u) + s\, e^{- \frac{u}{2 a(u)}} + \frac{\kappa u}{8 a^2(u)} + \cdots \ , 
\end{align}
whose time evolution is 
\begin{align}
 \dot R_s(u) 
 &= \dot a(u) - \frac{s}{2a(u)} e^{- \frac{u}{2a(u)}} + \frac{\kappa}{4a^2(u)} + \mathcal O(\kappa^2) 
\\
 &= - \frac{s}{2a(u)} e^{- \frac{u}{2a(u)}} + \mathcal O(\kappa^2) \ . 
\end{align}
Since $s>0$, the areal radius at the shell is always decreasing with time. 

%The following argument is wrong because the time lapse is long.
%Although the areal radius at the shell is still large, 
%it is expected that the proper distance below the apparent horizon 
%is very small as the Planck length
%(in the domain of approximation \eqref{near-horizon-region}),
%since the metric components of the 2 dimensional part of $(u,r)$, 
%is of $\mathcal O(\kappa)$. 
%We will estimate the proper distance in the next subsection. 

%By comparing with the position of the local minimum of $R$ \eqref{neck-VN}, 
%the collapsing shell is placed near (under) the local minimum of $R$
%at a distance of $\mathcal O(\kappa)$,
%\be
%\Delta r \simeq \frac{\kappa}{4a},
%\ee
%which is in exact agreement with the result \eqref{Deltar}
%obtained in the previous section.
%-10/10

%%%%%%%%%%%%%%%%%%%%%%%%%%%%%%%%%%%%%%%%%%%%%%%%%%%%%%%%%%%%%%%%%%%%%%%%%%%%%%%%
%%%%%%%%%%%%%%%%%%%%%%%%%%%%%%%%%%%%%%%%%%%%%%%%%%%%%%%%%%%%%%%%%%%%%%%%%%%%%%%%

\subsection{Distance Between Apparent Horizon and Collapsing Shell}
\label{sec:distance-pert}

In this subsection, we calculate more precisely the distance from the collapsing shell 
to the apparent horizon. 
In order to estimate the distance, 
we consider geodesics between 
the collapsing shell and the apparent horizon. 

In the previous sections, we have seen that 
%YM-10/10
the quantum corrections become important in $R$ 
but are negligible in $f$, in the metric \eqref{vaidya}. 
%\begin{equation}
% f(u,r) = 1 - \frac{a(u)}{r}  \ , 
%\end{equation}
%or equivalently, 
%\begin{equation}
% \tilde f(u,z) = 2 \dot a(u) + \frac{\kappa z}{a(u)} \ , 
%\end{equation}
So, $f$ is simply given by eq.\eqref{sol-f} %for $r - a = \mathcal O(\kappa)$ 
up to the corrections of $\mathcal O(\kappa^2)$. 
We expect that the higher-order corrections of $f$ would not give 
divergences around $r = a(u)$,
unlike the areal radius $R$. 
Thus, naive perturbation around the Vaidya metric implies that 
the expression \eqref{sol-f} would be good everywhere $r\geq a(u)$, 
although the perturbation breaks down due to the divergence in $R$. 
Here, we focus on the 2-dimensional part of $(u,r)$-directions. 
Since $R$ is irrelevant for this 2-dimensional part, 
we assume that eq.\eqref{sol-f} even holds deep inside the neck, 
and consider
%%% PM +
$r - a \leq \mathcal O(\kappa)$
%%% PM -
(including $r - a \ll \kappa/a$).%
\footnote{%
The divergence in $R$ may affect the validity of the expression \eqref{sol-f} deep inside the neck.
We will study the region deep inside the neck in the next section. 
The conclusion remains the same. 
} 
%-10/10

It is convenient to introduce a new coordinate $\chi$ as 
\begin{equation}
 r = a(u) - 2 a(u) \dot a(u) + \frac{\kappa}{4 a(u)} \chi^2 \ , 
\end{equation}
or equivalently, 
\begin{equation}
 z = - 2 \kappa^{-1} a(u) \dot a(u) + \frac{\chi^2}{4 a(u)} \ ,
\end{equation}
%%% PM +
with $\chi = \mathcal{O}(1)$.
%%% PM -
%YM-10/10
%-10/10
Since we assume that variation of the Schwarzschild radius with time is very slow
($\dot a(u) = \mathcal O(\kappa)$),
$a(u)$ can be treated as a constant
in the domain of the approximation \eqref{near-horizon-region}.

The (2-dimensional part of) the metric is expressed as 
%YM-10/10
\begin{equation}
 ds^2 = - \kappa \frac{\chi^2}{4 a^2} du^2 - \kappa \frac{\chi}{a} d\chi \, du \ . 
\end{equation}
%-10/10
We define the $\tau$-coordinate as 
\begin{align}
 du &= d\tau - 2 a \frac{d\chi}{\chi} \ , 
\end{align}
and the metric becomes
\begin{equation}
 ds^2 = \frac{\kappa}{4 a^2} \left( - \chi^2 d\tau^2 + 4a^2 d\chi^2\right) \ . \label{Rindler}
\end{equation}
This is nothing but the Rindler space. 

In the original $r$-coordinate,
the past horizon is located at 
\begin{align}
 r &= a(u) - 2 a(u) \dot a(u) \ .
\end{align}
%in the exterior geometry of the collapsing shell. 
%YM-10/10
The collapsing shell is hence always located outside the past horizon.% 
\footnote{%
%The geometry of the black-hole evaporation is constructed 
%by using the exterior geometry which we study here for the outside of the collapsing shell 
%and flat spacetime for the inside of the shell. 
%The exterior geometry is cut at the shell and interior is replaced by the flat space. 
%As the collapsing shell is always located outside the past horizon, 
%the past horizon does not appear in the black-hole geometry. 
The spacetime geometry \eqref{Rindler} under consideration only applies to
the exterior space of the collapsing shell.
(The interior space is flat spacetime.)
Hence the past horizon actually does not exist.
}
%-10/10
%This past horizon is identical to that of the Rindler space, \eqref{Rindler}. 
The apparent horizon is located at 
\begin{equation}
 r = a(u) - 2 a(u) \dot a(u) + \frac{\kappa}{4 a(u)} \ , 
\end{equation}
which corresponds to $\chi = 1$. 

To calculate the proper distance, 
it is convenient to use the coordinates of the flat spacetime; 
\begin{align}
 T &= \chi \sinh(\tau/2a) \ , 
 &
 X &= \chi \cosh(\tau/2a) \ , 
\end{align}
in terms of which the metric is
\begin{equation}
 ds^2 = \kappa \left( - dT^2 + dX^2\right) \ . 
\end{equation}

The apparent horizon, or equivalently the $\chi=1$ line,
is a hyperbolic curve $-T^2 + X^2 = 1$. 
The distance between the origin $X=T=0$ to the apparent horizon $\chi=1$ 
is always 1 in the Minkowski space up to the factor $\sqrt{\frac{\kappa}{4a^2}}$. 
Since the collapsing shell is located on the future side of the past horizon $X + T = 0$, 
the distance from any point on the collapsing shell to the apparent horizon
before the shell crosses the future horizon 
is always smaller than that from the origin. 
Thus the distance is smaller than $\ell_p \equiv \kappa^{1/2}$.
%except for the final stage of the evaporation $a = \mathcal O(\ell_p)$. 
%which is of $\mathcal O(\ell_p)$ as long as the Schwarzschild radius 
%is sufficiently larger than the Planck length, $a = \mathcal O(\kappa^0)$.

The calculation above about the Planckian distance
between the collapsing shell and the apparent horizon 
is based on the metric derived for the near-horizon region.
(Furthermore, we have ignored the time-dependence by dropping $\dot{a}$.)
A priori it does not have to hold when the shell is deep inside the horizon.
More discussions on the scale of this proper distance deep inside the horizon
will be given in the next section.

%%%%%%%%%%%%%%%%%%%%%%%%%%%%%%%%%%%%%%%%%%%%%%%%%%%%%%%%%%%%%%%%%%%%%%%%%%%%%%%%
%%%%%%%%%%%%%%%%%%%%%%%%%%%%%%%%%%%%%%%%%%%%%%%%%%%%%%%%%%%%%%%%%%%%%%%%%%%%%%%%
%%%%%%%%%%%%%%%%%%%%%%%%%%%%%%%%%%%%%%%%%%%%%%%%%%%%%%%%%%%%%%%%%%%%%%%%%%%%%%%%

\section{Deep Inside The Neck}
\label{sec:deep-dyn}

So far, we have studied the spacetime geometry
from the asymptotic region to the near-horizon region,
including a small neighborhood slightly inside the apparent horizon
where $R-a = \mathcal O(\kappa/a)$.
In this section,
we shall study the geometry deeper inside the apparent horizon
where $R-a = \mathcal O(a)$.

%%% 0406-1
Note that
%%% 0406-2
the phrase ``deep inside the neck'' in this paper %the title of this section
does not imply a large proper distance in the radial direction
between the apparent horizon and a point in this region.
It merely refers to a region inside the neck
where the areal radius $R$ is significantly larger than its value at the neck.

%As we have discussed in the previous sections,
%the areal radius $R_s$ on the collapsing shell is always 
%decreasing with the retarded time $u$,
%and $R_s(u)$ becomes much smaller than $a(u)$ 
%since the Schwarzschild radius $a(u)$ is decreasing faster than $R_s(u)$ 
%because of the Hawking radiation. 
%Thus, in order to study the evaporation of the black hole, 
%we need to study the region deeper inside the neck where $R-a\sim\mathcal O(a)$. 

\subsection{Dynamical Geometry Deep Inside The Neck}

%While the static solution in this region is already given in Sec.~\ref{sec:deep-stat}, 
%the time-dependent solution is needed
%to study the dynamical process of the collapse. 
In this subsection,
we study the dynamical geometry deep inside the neck
in the sense that $R - a$ is of order $\mathcal{O}(\kappa^0)$,
%YM-11/12
for arbitrary %the generic situation when both 
$\beta$ and $\gamma$ in the semi-classical Einstein equations
\eqref{Tuu-vac} and \eqref{Tvv-vac}. % are non-zero.
%-11/12
We will determine the values of $\beta$ and $\gamma$ in the next subsection
for a collapsing null shell.

According to eq.\eqref{C-neck},
the red-shift factor $C$ becomes exponentially small 
when $x_0 - x \gg a(u)$.
%$C\sim \exp\langle\mathcal O(\kappa^{-1})\rangle]$ 
Therefore,
in the region deep inside the neck where eq.\eqref{x-x0} holds,
%YM-11/12
%we expect that one can ignore 
higher-order terms in the $C$-expansion can be ignored. 
%(The validity of this assumption can be verified in the end of the calculation.)
%-11/12
With this approximation,
the semi-classical Einstein equation is approximated by
the following equations:
\begin{align}
 0 
 &= 
 - 2 \kappa \beta R^2 + \left(\partial_u R^2\right)^2 
 + 4 R^2 \left(\partial_u R^2\right) \partial_u \rho 
 - 2 R^2 \left[\left(\partial_u R^2\right)^2 
 - 2 \kappa \left(\partial_u \rho\right)^2 + 2 \kappa \partial_u^2 \rho\right] \ , 
 \label{equu-deep}
\\
 0 
 &= 
 - 2 \kappa \gamma R^2 + \left(\partial_v R^2\right)^2 
 + 4 R^2 \left(\partial_v R^2\right) \partial_v \rho 
 - 2 R^2 \left[\left(\partial_v R^2\right)^2 
 - 2 \kappa \left(\partial_v \rho\right)^2 + 2 \kappa \partial_v^2 \rho\right] \ , 
 \label{eqvv-deep}
\\
 0 
 &\simeq
 \partial_u \partial_v R^2 + 2 \kappa \partial_u \partial_v \rho \ . 
 \label{equv-deep}
\end{align}
The first two equations \eqref{equu-deep} and \eqref{eqvv-deep}
are exactly equivalent to $G_{uu} = \kappa \langle T_{uu}\rangle$
and $G_{vv} = \kappa \langle T_{vv}\rangle$.
%YM-11/12
The third equation \eqref{equv-deep} is 
equivalent to $G_{uv} = \kappa \langle T_{uv}\rangle$
up to the term $C/2$,
which can be neglected for small $C$. %in the right-hand side of the equation.
%-11/12

From eq.\eqref{equv-deep}, we obtain 
\begin{equation}
 \rho \simeq - \frac{R^2}{2\kappa} + F(v) + \bar F(u) \ , 
\label{sol-rho-deep}
\end{equation}
where $F(v)$ and $\bar F(u)$ are the integration constants. 

Remarkably,
the other two nonlinear differential equations \eqref{equu-deep}, \eqref{eqvv-deep}
%YM-11/12
can be solved exactly (to all orders in $\kappa$) by using eq.\eqref{sol-rho-deep} as 
%-11/12
\begin{equation}
 \sqrt{R^2(R^2 - \kappa)} 
 - \alpha \log\left(R + \sqrt{R^2 - \kappa}\right)
 = 
 G(v) + \bar G(u) \ . 
\label{sol-r-deep}
\end{equation}
where $G(v)$ and $\bar G(u)$ are given by 
\begin{align}
 G(v) 
 &= 
 \pm 2 \kappa \int^v dv' \sqrt{F^{\prime\,2}(v') - F''(v') -  \frac{1}{2} \gamma(v')} \ , 
\label{sol-gv-deep}
\\
 \bar G(u) 
 &= 
 \pm 2 \kappa \int^u du' \sqrt{\bar F^{\prime\,2}(u') - \bar F''(u') - \frac{1}{2} \beta(u')} \ .  
\label{sol-gu-deep}
\end{align}

Since we are considering the geometry deep inside the neck, 
where the areal radius is significantly larger than the Schwarzschild radius $a$,
%Whereever the areal radius $R$ is not very small
%but of order $\mathcal O(\kappa^0)$, 
eq.\eqref{sol-r-deep} is approximated by 
\begin{align}
 R^2 &= G(v) + \bar G(u) - \kappa \log R + \mathcal O(\kappa^2) \ , 
\notag\\
 &= G(v) + \bar G(u) - \frac{\kappa}{2} \log\left(G(v) + \bar G(u)\right) + \mathcal O(\kappa^2) \ , 
\label{r-approx-deep}
\end{align}
where some constants are absorbed by redefining $G(v)$ and $\bar G(u)$. 
Using eq.\eqref{sol-rho-deep},
$\rho$ is given by
\begin{equation}
 \rho = F(v) + \bar F(v) - \frac{1}{2\kappa} \left(G(v) + \bar G(u)\right) - \frac{1}{2} \log R + \mathcal O(\kappa) \ . 
\label{rho-approx-deep}
\end{equation}

The integration constants $F(v)$ and $\bar F(u)$ are related to 
the choice of the null coordinates $u$ and $v$. 
We consider the coordinate transformation to $(U,V)$-coordinates 
for which the metric is expressed as 
\begin{equation}
 ds^2 = \widetilde C dU dV + R^2 d \Omega^2 \ , 
\end{equation}
where $\widetilde C$ is related to $C$ via
\begin{equation}
 \widetilde C = \frac{\partial u}{\partial U} \frac{\partial v}{\partial V} C \ . 
\end{equation}
Then, $\rho$ transforms to $\tilde \rho = \frac{1}{2} \log \widetilde C$ as 
\begin{equation}
 \tilde \rho = \rho 
 - \frac{1}{2}\log\left(\frac{dU}{du}\right) 
 - \frac{1}{2}\log\left(\frac{dV}{dv}\right) \ . 
\end{equation}
The solution for $\rho$ is expressed as 
\begin{align}
 \tilde \rho = - \frac{R^2}{2\kappa} + \widetilde F(V) + \bar {\widetilde F}(U) \ , 
\end{align}
where the integration constants $F(v)$ and $\bar F(u)$ are transformed to 
$\widetilde F(V)$ and $\bar {\widetilde F}(U)$ as 
\begin{align}
 F(v) &= \widetilde F(V) + \frac{1}{2}\log\left(\frac{dV}{dv}\right) \ , 
 \label{trans-F}
\\
 \bar F(u) &= \bar{\widetilde F}(U) - \frac{1}{2}\log\left(\frac{dU}{du}\right) \ . 
 \label{trans-Fbar}
\end{align}

The solution for $R^2$ is given by the same equation \eqref{sol-r-deep} 
but now $G$ and $\bar G$ should be replaced by those for the new coordinates $U$ and $V$ as
\begin{align}
 \widetilde G(V)
 &= 
 \pm 2 \kappa \int^V dV' 
 \sqrt{\widetilde F^{\prime\,2}(V') - \widetilde F''(V') - \frac{1}{2} \tilde \gamma(V') } \ , 
 \label{Gtilde}
\\
 \bar {\widetilde G}(U)
 &= 
 \pm 2 \kappa \int^U dU' 
 \sqrt{\bar{\widetilde F}^{\prime\,2}(U') 
   - \bar{\widetilde F}''(U') - \frac{1}{2} \tilde \beta(U')} \ . 
   \label{Gtildebar}
\end{align}

Since the first terms in the expressions
\eqref{Tuu-vac} and \eqref{Tvv-vac} are not covariant,
$\tilde \beta$ and $\tilde \gamma$ receive 
corrections under the coordinate transformation as 
\begin{align}
 \beta \to \tilde \beta 
 &= 
 \left(\frac{du}{dU}\right)^2 \left(\beta + \frac{1}{2} \{U,u\}\right) \ , 
 \label{trans-beta}
\\
 \gamma \to \tilde \gamma 
 &= 
 \left(\frac{dv}{dV}\right)^2 \left(\gamma + \frac{1}{2} \{V,v\}\right) \ . 
 \label{trans-gamma}
\end{align}
where $\{f,x\}$ is the Schwarzian derivative of $f$ with respect to $x$. 
By using eqs.\eqref{trans-F}, \eqref{trans-Fbar}, \eqref{trans-beta} and \eqref{trans-gamma}, 
it is straightforward to see that,
according to their definitions \eqref{Gtilde}, \eqref{Gtildebar},
\eqref{sol-gv-deep} and \eqref{sol-gu-deep}, 
\begin{align}
 G(v) &= \widetilde G(V) \ , 
 &
 \bar G(u) &=  \bar {\widetilde G}(U) \ . 
\end{align}

%YM-10/10
%%%%%%%%%%%%%%%%%%%%%%%%%%%%%%%%%%%%%%%%%%%%%%%%%%%%%%%%%%%%%%%%%%%%%%%%%%%%%%%%
%%%%%%%%%%%%%%%%%%%%%%%%%%%%%%%%%%%%%%%%%%%%%%%%%%%%%%%%%%%%%%%%%%%%%%%%%%%%%%%%

\subsection{Junction Condition}
\label{sec:junction-deep}

The solution \eqref{sol-rho-deep}-\eqref{sol-gu-deep}
of the semi-classical Einstein equation for the region deep inside the neck
is given in terms of two arbitrary functions $F(v)$ and $\bar F(u)$, 
which are related to the definition of the coordinates $v$ and $u$, respectively. 
They can be determined by fixing the coordinates. 
The relation between the coordinate in this deep region and that in the outer region 
is determined by the junction condition of the metric. 
In this subsection,
we shall choose the same coordinates $u$ and $v$ in the asymptotically flat region, 
which is defined as $u=t-r$ and $v=t+r$ in the asymptotic Minkowski space. 
Near the Schwarzschild radius, they are related to the tortoise coordinate $x$ 
in Sec.~\ref{sec:review}--\ref{sec:neck-shell} via \eqref{x2uv}. 

%%%%%%%%%%%%%%%%%%%%%%%%%%%%%%%%%%%%%%%%%%%%%%%%%%%%%%%%%%%%%%%%%%%%%%%%%%%%%%%%

\subsubsection{Static Case}
\label{sssec:junction-deep-stat}

%In the rest of this section,
We first consider the static case and 
check that the static solution in Sec.~\ref{sec:deep-stat}
is a special case of the general dynamical solution obtained above.
%For the static geometry inside the local minimum of the radius $R$, 
%$R$ increases along the incoming null lines and decreases along the outgoing null lines, 
%namely, 
%\begin{align}
% \left( \frac{\partial R^2}{\partial v}\right)_u &\simeq G'(v) < 0 \ , 
%& 
% \left( \frac{\partial R^2}{\partial u}\right)_v &\simeq \bar G'(u) > 0 \ , 
% \label{SignGstat}
%\end{align}
%and the signs in \eqref{sol-gv-deep} and \eqref{sol-gu-deep} 
%should be chosen by these conditions. 
%
The static solution is independent of the time $t$,
and the null coordinates can be chosen such that $u = t - x$ and $v = t + x$. 
%$R^2 = G(v) + \bar G(u)$ is independent of $t$ and 
%so is $F(v) + \bar F(u)$, implying that 
Since $F(v) + \bar F(u)$ is independent of $t$, we have
\begin{align}
 F(v) &= k_0 + \frac{1}{4\kappa} k_1 v \ , 
 \label{F-static}
 &
 \bar F(u) &= \bar k_0 - \frac{1}{4\kappa} k_1 u \ , 
\end{align}
where $k_0$, $\bar{k}_0$ and $k_1$ are constants. 
%For the static case, by identifying the null coordinates with those in the asymptotically flat region, 
Since the null coordinates $u$ and $v$ are chosen such that 
they are identical to those in the asymptotically flat region, 
we have $\beta(u) = \gamma(v) = 0$
for the static solution. 
Substituting this static condition \eqref{F-static} to 
eqs.\eqref{sol-rho-deep}--\eqref{sol-gu-deep} %and \eqref{r-approx-deep}, 
the static solution is obtained as 
\begin{align}
 \rho &= k_0 + \bar k_0 + \kappa^{-1} k_1 x \ , 
\\
 R^2 &= k_0 + \bar k_0 + k_2 - k_1 x \ , 
\end{align}
where the signs in eqs.\eqref{sol-gv-deep} and \eqref{sol-gu-deep} 
are chosen such that $R^2$ is independent of $t$,
and that $\rho$ is non-trivial.%
\footnote{%
There is another solution for which $\rho$ is a constant at the leading order. 
} 
This is consistent with eqs.\eqref{rho-x-deep-stat} and \eqref{r-x-deep-stat}. 
After patching the geometry with that near the neck, 
the constants above are fixed as 
\begin{align}
 k_0 + \bar k_0 &\simeq - \frac{x_0}{2a} \ , &
 k_1 &\simeq \frac{\kappa}{2a} \ , 
 &
 k_2 &\simeq a^2 \ . 
\end{align}

%%%%%%%%%%%%%%%%%%%%%%%%%%%%%%%%%%%%%%%%%%%%%%%%%%%%%%%%%%%%%%%%%%%%%%%%%%%%%%%%

\subsubsection{Dynamical Case}
\label{sssec:junction-deep-dyn}

Now, we look for explicit expressions of the metric functions $\rho$ and $R$
for the dynamical case
through eqs.\eqref{sol-gv-deep}, \eqref{sol-gu-deep}, \eqref{r-approx-deep}, and \eqref{rho-approx-deep}
by investigating the junction condition.

In the previous subsection,
we have chosen the signs in eqs.\eqref{sol-gv-deep} and \eqref{sol-gu-deep} 
by using the static condition of $R^2$ and the condition of the non-trivial behavior of $\rho$. 
However, it is not obvious that $\rho$ should not be a constant, 
and the signs should be chosen by the junction condition near the neck. 
For the static case, 
%since $\beta = \gamma = 0$, 
%YM: This is incorrect since \beta and \gamma depend on the coordinates u and v. 
%    but the structure explained below is independent of the gauge. 
the radius $R$ inside the neck increases along incoming null lines 
and decreases along outgoing null lines; 
\begin{align}
 \left( \frac{\partial R^2}{\partial v}\right)_u &\simeq G'(v) < 0 \ , 
& 
 \left( \frac{\partial R^2}{\partial u}\right)_v &\simeq \bar G'(u) > 0 \ , 
 \label{SignGstat}
\end{align}
and the signs in eqs.\eqref{sol-gv-deep} and \eqref{sol-gu-deep} 
are consistent with this condition. 
On the other hand, 
%when we consider the collapsing shell,
the geometry for the collapsing shell is constructed by connecting the dynamical solution
\eqref{sol-rho-deep}--\eqref{sol-gu-deep} 
to the flat spacetime through the collapsing shell. 
The junction condition implies that $R$ on the shell 
in the geometry \eqref{sol-rho-deep}-\eqref{sol-gu-deep} 
must be consistent with that in the flat spacetime. 
As the radius must decrease along the incoming null line 
in the flat spacetime, $R$ must also decrease 
in the solution \eqref{sol-rho-deep}-\eqref{sol-gu-deep}. 
Hence, we have the conditions
\begin{align}
 \left( \frac{\partial R^2}{\partial v}\right)_u &\simeq G'(v) < 0 \ , 
& 
 \left( \frac{\partial R^2}{\partial u}\right)_v &\simeq \bar G'(u) < 0 \ , 
\end{align}
and both signs in eqs.\eqref{sol-gv-deep} and \eqref{sol-gu-deep} must be minus. 

Now, we consider the junction condition to the near-neck region. 
Near the neck where
$R - a = \mathcal O(\kappa)$,
using eqs.\eqref{C-neck}, \eqref{x2uv},
we have
\begin{equation}
 \rho = \frac{v-u}{4a(u)} + \mbox{const.} + \mathcal{O}(\kappa) \ .
 %- \frac{1}{2}\log \frac{a^2}{\kappa} \ , 
\label{rho-neck-uv}
\end{equation}
%where we have shifted the coordinates $u$ and $v$
%$u-v\simeq 0$ 
%such that $C(0, 0) = e^{2\rho(0, 0)} = \kappa/a^2$.
%The point $(u = 0, v = 0)$ can be identified
%with a point around the neck,
%where $C \sim \mathcal{O}(\kappa/a^2)$
%according to the Schwarzschild metric.
This expression is valid near the neck where $R-a = \mathcal O(\kappa)$,
and can be extended to the slightly deeper region
with $R - a \sim - \kappa \log \kappa$,
where $C \sim \mathcal O(\kappa^2)$.  

Using the expression \eqref{rho-approx-deep}, 
$\rho$ has the following general form deep inside the neck
for $R - a \sim \mathcal{O}(a)$:
\begin{equation}
 \rho = W(v) + \bar W(u) - \frac{1}{2}\log R + \mathcal O(\kappa) \ . 
\label{rho-w}
\end{equation}
This expression is consistent with eq.\eqref{rho-neck-uv} around the neck.
Hence it can be continuously patched with the solution \eqref{rho-neck-uv}
over the region from $R - a \sim \mathcal O(\kappa)$
to $R - a \sim - \kappa \log \kappa$.
The patching requires $W(v)$ and $\bar W(u)$ to be approximately given by
\begin{align}
 W(v) &\simeq \int^v \frac{dv'}{4\tilde a(v')} \ , 
%\label{W(v)}
 & 
 \bar W(u) &\simeq - \int^u \frac{du'}{4a(u')} \ , 
%\label{W(u)}
\label{w}
\end{align}
%YM-11/12
where $\tilde a(v) = a(u_A(v))$ and $u_A(v)$ is 
%-11/12
the $u$-coordinate of the apparent horizon for a given value of $v$.%
%%% PM +
%[Moving this footnote here from below.]
\footnote{%
It should be noted that $a(u) = a(u_A(v))$ only at the apparent horizon. 
%The functions $a(u)$ and $a(v)$ stand for the apparent horizon at $u$ and $v$, respectively. 
%In the deeper region, they are different for given $u$ and $v$; $a(u) \neq a(v)$. 
}
%%% PM -
Since the apparent horizon is located at $x \equiv \frac{1}{2}(v-u) = x_A$, 
$u_A(v)$ is given by 
\begin{equation}
 u_A(v) = v - 2 x_A(v) \ , 
\end{equation}
%%% PM +
% [I think it is better to keep the notation more explicit.]
%YM-11/12
As the time evolution of $x_A$ is of $\mathcal O(\kappa)$, 
$u$- or $v$-dependence of $x_A$ can be treated as a higher-order correction. 
%-11/12
%Hereafter, we simply refer to $a(u_A(v))$ as $a(v)$.%
%%% PM - 
%$a(u)$ and $\bar{a}(v)$ are defined such that $a(u) = \bar{a}(v)$ at the neck. 
%Since $a(u)$ is a function of $u$ and independent of $x = \frac{1}{2}(v-u)$, 
%$v$ dependence of $a$ is given by the same function. 
%
We have used the integration in eq.\eqref{w} so that
eq.\eqref{rho-w} agrees with eq.\eqref{rho-neck-uv}
over a short period of time of $\mathcal{O}(a)$
in which $a(u)$ is almost a constant up to $\mathcal{O}(\kappa)$.
After the patching is done,
the expression \eqref{rho-w} holds everywhere inside the neck
in vacuum where $C$ is sufficiently small.
%Since $W(v)$ is independent of $u$, 
%$v$-dependence in the deeper region is same as that near the neck \eqref{w}, 
%for same $v$ (but different $u$), and similarly for $\bar W(u)$. 

We impose the conditions \eqref{rho-w}--\eqref{w} to the solution 
\eqref{sol-rho-deep}--\eqref{sol-gu-deep}. 
For $F(v)$, we impose the initial condition 
that there is no incoming energy in the past null infinity, $\gamma=0$. 
For $\bar F(u)$, we impose the junction condition on the shell 
that the outgoing energy is continuous at the shell and hence $\langle T_{uu}\rangle = 0$. 
Together with these boundary conditions, 
we obtain 
\begin{align}
 F(v) &\simeq \int^v \frac{dv'}{8 \tilde a(v')} \ , 
 & 
 \bar F(u) &\simeq - \int^u \frac{du'}{4 a(u')} \ .  
\end{align}
$G(v)$ and $\bar G(u)$ are calculated as 
\begin{align}
 G'(v)
 &\simeq 
 - 2 \kappa F'(v) 
 \simeq - \frac{\kappa}{4 \tilde a(v)}
 \ , 
 \\
 \bar G'(u)
 &\simeq 
 - 2 \kappa \sqrt{- \bar F''(u)} 
%\\
% &
 \simeq 
 - 2 \kappa \sqrt{- \frac{\dot a(u)}{4 a^2(u)}} = \mathcal O(\kappa^2) \ .
% \simeq 
% - \sqrt{ \frac{\alpha^3}{8 a^4(u)}} \ .  
\label{G'(u)}
\end{align}
Eq.~\eqref{G'(u)} implies that the areal radius $R$ is almost constant 
along the null lines of $v = \text{const}$. 
By using the expressions above, 
it is straightforward to see that the formula for the Hawking radiation \eqref{da/du-neck-2} 
is still valid even if the collapsing shell is deep inside the neck. 
Thus, we obtain 
\begin{equation}
 \bar F(u) \simeq \int^u 2 \kappa^{-1} a(u') \dot a(u') du' = \kappa^{-1} a^2(u) \ , 
\end{equation}
and, similarly for $F(v)$ and $G(v)$, 
\begin{equation}
 G(v) = 2 \kappa F(v) + \text{const.} \simeq \tilde a^2(v) \ . 
\end{equation}
Thus, $\rho$ and $R$ are expressed as 
\begin{align}
 \rho 
 &= 
 \kappa^{-1} \left(a^2(u) - \tilde a^2(v)\right) 
 + \frac{\tilde x_A(u) - x_0(u)}{2 a(u)} - \frac{1}{2}\log \tilde a(v) 
 + \mathcal O(\kappa) \ , 
 \label{rho-explicit}
 \\
 R
 &= 
 \tilde a(v) + \mathcal O(\kappa) \ , 
\end{align}
where $\tilde x_A(u)$ is defined by the junction condition such that 
eq.\eqref{rho-explicit} is consistent with $C$ \eqref{C-neck} (with eq.\eqref{xapp})
at the apparent horizon,
%\footnote{%
%Note that $a(u) = a(u_A(v))$ at the apparent horizon $x = x_A(u)$. 
%} 
and it is given by 
\begin{equation}
 \tilde x_A(u) %= x_A(u) + a(u) \log\left(\frac{a(u)}{4}\right) 
 = x_0(u) - a(u) \log\left(\frac{4a(u)}{\kappa}\right) \ . 
\end{equation}
Hence, $C$ is now obtained as
\footnote{
Note that $a^2(u) - a^2(u_A(v)) < 0$
for $(u, v)$ inside the apparent horizon.
}
\begin{equation}
 C \simeq \frac{\kappa}{4 a(u) \tilde a(v)} e^{2 (a^2(u) - \tilde a^2(v))/\kappa} \ . 
 \label{C-deep}
\end{equation}

The scalar curvature $\mathcal R$, squares of the Ricci tensor $\mathcal R_{\mu\nu}$ 
and Riemann tensor $\mathcal R_{\mu\nu \rho \sigma}$ 
%$\mathcal R_{\mu\nu} \mathcal R^{\mu\nu}$ and 
%$\mathcal R_{\mu\nu \rho \sigma} \mathcal R^{\mu\nu \rho \sigma}$, 
are estimated as 
\begin{align}
 \mathcal R 
 &\simeq 
 \frac{2}{\tilde a^2(v)} \ , 
&
 \mathcal R_{\mu\nu} \mathcal R^{\mu\nu} 
 &\simeq 
 \frac{2}{\tilde a^4(v)} \ , 
&
 \mathcal R_{\mu\nu \rho \sigma} \mathcal R^{\mu\nu \rho \sigma} 
 &\simeq 
 \frac{4}{\tilde a^4(v)} \ . 
\end{align}
The expressions of the curvature invariants above are valid 
deep inside the apparent horizon. 
For a given value of $v$ outside the collapsing shell,
as long as $\tilde a(v) = a(u_A(v))$ is large,
the curvature is small for any $u > u_A(v)$,
even when $a(u)$ is approaching the Planck scale.

%Eq.\eqref{r-approx-deep} gives
%\begin{align}
%R^2(u, v) \simeq& R^2(u_1, v_1)
%- \kappa\log\left(\frac{R(u, v)}{R(u_1, v_1)}\right)
%\nn \\
%&- \kappa \int_{u_1}^u du' \, \frac{1}{4a(u')}e^{-\int_{u_1}^u \frac{du''}{2a(u'')}}
%- \kappa \int_{v_1}^v dv' \, \frac{1}{4\bar{a}(v')}e^{\int_{v_1}^v \frac{dv''}{2\bar{a}(v'')}}
%+ \mathcal{O}(\kappa^{3/2}) \ ,
%\end{align}
%for an arbitrary reference point $(u_1, v_1)$
%inside the neck.
%{\color{blue}
%(Please check the formula above.)
%}

%%%%%%%%%%%%%%%%%%%%%%%%%%%%%%%%%%%%%%%%%%%%%%%%%%%%%%%%%%%%%%%%%%%%%%%%%%%%%%%%
%%%%%%%%%%%%%%%%%%%%%%%%%%%%%%%%%%%%%%%%%%%%%%%%%%%%%%%%%%%%%%%%%%%%%%%%%%%%%%%%

\subsection{Collapsing Shell Deep Inside The Neck}
\label{sec:deep-shell}

In the previous section, we have seen that 
the areal radius is almost constant along the incoming null lines, $v = \text{const}$. 
Although $R$ is decreasing along the collapsing shell, 
its decreasing rate is of $\mathcal O(\kappa^2)$ and 
the Schwarzschild radius $a(u)$ is decreasing much faster. 
%Thus the difference of the areal radius at the shell
%and that at the apparent horizon can be large. 
%If the areal radius of the shell is as small as the Planck scale 
%for a given $u=\text{const.}$ line where $a(u)$ is as small as the Planck scale, 
%interior of the neck will be very small and the information may escape if 
%physics of the Planck scale is taken into account. 
%YM-11/12
%In the conventional model,
This situation continues
as long as $a(u)$ is sufficiently larger than the Planck length.
Hence, when $a(u)$ becomes much smaller
than its initial value (but still larger than the Planck length),
the areal radius of the shell is still large at the same time $u$,
so the interior space inside the neck
(including the space occupied by the collapsing matter) is still large.
Naively, 
this large interior space inside the neck
either becomes disconnected from the outside world
as the neck shrinks to $0$ eventually,
resulting in the event horizon,
or they remain connected to the outside world
through tiny necks as remnants.
%-11/12

However, the expression \eqref{G'(u)} implies 
that the decreasing rate of the areal radius along the incoming null lines 
can be large when the Schwarzschild radius $a$ approaches the Planck length $\ell_p$. 
The semi-classical approximation will be invalid,
and the outgoing energy of the Hawking radiation $(\sim 1/a^2)$
will be at the Planck scale. 
Although the expression \eqref{G'(u)} would not be reliable when $a \sim \ell_p$, 
there would not be the problem of the cutoff scale 
if we focus on the locus of the collapsing shell. 
When the areal radius starts decreasing again, 
it is still much larger than the Planck length and comparable 
to the Schwarzschild radius for the initial mass of the shell. 
The outgoing energy $T_{uu}$ is also zero just on the shell. 
Thus it would be possible to study the behavior of the shell 
without referring to $a(u)$. 

In this subsection, we check that the areal radius at the shell starts to decrease again. 
Note that the new coordinate for $u$-direction at the shell cannot be 
connected to that in the asymptotic region when $a(u) = \mathcal O(\kappa)$, 
since the semi-classical approximation would not be valid around the apparent horizon in between. 

%-10/10

By choosing a suitable coordinate system $(\tilde{u}, \tilde{v})$,
we can set $\tilde{F}(\tilde{v}) = 0$ and $\bar{\tilde{F}}(\tilde{u}) = 0$,
so that the solution is simply given by (for $R^2\gg\mathcal O(\alpha)$)
\begin{align}
 \rho &\simeq - \frac{R^2}{2 \kappa} \ , 
\label{sol-rho1}
\\
 R^2 &\simeq G(\tilde v) + \bar G(\tilde u) \ , 
\label{sol-r1}
\end{align}
where
\begin{align}
 G(\tilde v) &= - 2 \kappa \int^{\tilde v} dv' \sqrt{-\frac{1}{2}\tilde \gamma(v')}
\label{sol-gv1}
\\
 \bar G(\tilde u) &= - 2 \kappa \int^{\tilde u} du' \sqrt{-\frac{1}{2}\tilde \beta(u')} \ , 
\label{sol-gu1}
\end{align}

Note that the coordinates $\tilde u$ and $\tilde v$
in the expressions above are different 
from the original null coordinates, 
and $\tilde \gamma(\tilde v)$ and $\tilde \beta(\tilde u)$ are different 
from the incoming and outgoing radiation
in the asymptotically flat past and future null infinities
(see eqs.\eqref{trans-beta}, \eqref{trans-gamma}). 
The outgoing energy-momentum tensor is calculated as 
\begin{equation}
 R^2 \langle T_{uu}\rangle 
 = 2 \tilde \beta(\tilde u) + \left(- 2 \tilde \beta(\tilde u)\right)^{-1/2} \dot{\tilde \beta}(\tilde u) \ . 
\end{equation}
Since $\langle T_{uu}\rangle$ must be continuous across the incoming null shell, 
we impose the junction condition $\langle T_{uu}\rangle = 0$,
and then, $\tilde \beta(\tilde u)$ is solved as 
\begin{equation}
 \tilde \beta = - \frac{1}{2 (\tilde u - \tilde u_0)^2} \ .
\end{equation}
The solution for $\beta$ is negative,
so that $\bar G(\tilde u)$ is real. 
$\bar{G}'(\tilde u)$ is monotonically increasing or decreasing depending on 
whether $\tilde u > \tilde u_0$ or $\tilde u < \tilde u_0$. 
By comparing with \eqref{G'(u)}, $\bar G'(\tilde u)$ should decrease with time. 
In this case, the areal radius $R$ on the shell behaves as 
\begin{equation}
 R^2 = \bar G(\tilde u) + \text{const.} \simeq \kappa\log(\tilde u_0-\tilde u) + \text{const.} \ , 
\end{equation}
which goes to zero at some point, $\tilde u = \tilde u_e < \tilde u_0$. 
Therefore, the collapsing shell eventually reaches $R=0$ in the exterior geometry.

\subsection{Distance Between Apparent Horizon and Collapsing Shell}
\label{sec:distance-deep}

In this subsection, 
using the explicit metric derived above deep inside the horizon,
we will show that the proper distance from the shell to the apparent horizon is indeed very short
even when the collapsing shell has moved to the deeper region.

Now, we calculate the proper distance between the collapsing shell and the apparent horizon. 
Using the explicit form of \eqref{C-deep}, 
we can see that the $(u,v)$-subspace is 2D flat spacetime:
\begin{equation}
 ds^2 = \frac{\kappa}{4a(u) \tilde a(v)} e^{2(a^2(u) - \tilde a^2(v))/\kappa} du\,dv  = \kappa dU\,dV \ , 
 \label{metric-UV-1}
\end{equation}
where the new coordinates $U$ and $V$ are defined as 
\begin{align}
 U(u) 
 &= 
 - \int_u^{u_c} \frac{1}{2a(u')} e^{2a^2(u')/\kappa} du' \ , 
 &
 V(v) 
 &= 
 \int_{v_s}^v \frac{1}{2\tilde a(v)} e^{-2\tilde a^2(v')/\kappa} dv' \ , 
\end{align}
where $v=v_s$ is the position of the collapsing null shell, 
and $u=u_c$ is the retarded time at the evaporation.%
\footnote{%
To be more precise, $u_c$ should be the cutoff time when 
the Schwarzschild radius $a(u)$ becomes sufficiently small 
but still larger than the Planck scale. 
} 
Recall that $a(u)$ and $\tilde a(v) = a(u_A(v))$ stand for the areal radius of the apparent horizon 
at $u$ and $v$, respectively, and hence, satisfy 
$a^2(u) - \tilde a^2(v) \leq 0$ since $a$ is decreasing with time. 
By using the formula of the Hawking radiation \eqref{da/du-neck-2}, 
$U$ and $V$ are expressed as 
\begin{align}
 U(u) 
 &= 
 \left(1 - e^{2a^2(u)/\kappa}\right) \ , 
 &
 V(v) 
 &= 
 \left(e^{-2\tilde a^2(v)/\kappa} - e^{-2a_0^2/\kappa}\right) \ , 
\end{align}
where $a_0$ is the Schwarzschild radius for the initial mass of the shell, 
or equivalently, the maximum of $a$. 

Consider two arbitrary points $(u, v)$ and $(u', v')$ inside the apparent horizon
with a spacelike separation.
Their proper distance $L$ is bounded from above by
\begin{align}
 L^2 
 &= 
 - \kappa (U(u) - U(u')) (V(v) - V(v')) 
\notag\\
 &< 
 \kappa \left(e^{2a^2(u)/\kappa} - e^{2a^2(u')/\kappa}\right) 
 \left(e^{-2\tilde a^2(v)/\kappa} - e^{-2\tilde a^2(v')/\kappa}\right) 
\notag\\
 &< 
 \kappa \, e^{2(a^2(u) - \tilde a^2(v))/\kappa} \leq \kappa = \ell_p^2 \ . 
\end{align}
(Without loss of generality,
we assume that $(u', v')$ is deeper inside the horizon than $(u, v)$,
i.e. $u'>u$ and $v'<v$.)
Therefore, the proper distance between two arbitrary points 
inside the apparent horizon is of $\mathcal O(\ell_p)$.

%%%%%%%%%%%%%%%%%%%%%%%%%%%%%%%%%%%%%%%%%%%%%%%%%%%%%%%%%%%%%%%%%%%%%%%%%%%%%%%%
%%%%%%%%%%%%%%%%%%%%%%%%%%%%%%%%%%%%%%%%%%%%%%%%%%%%%%%%%%%%%%%%%%%%%%%%%%%%%%%%
%%%%%%%%%%%%%%%%%%%%%%%%%%%%%%%%%%%%%%%%%%%%%%%%%%%%%%%%%%%%%%%%%%%%%%%%%%%%%%%%

\section{Conclusion and Discussion}\label{sec:conclusion}

In this paper, we have studied effects of the vacuum energy-momentum tensor
in the formation and evaporation of a black hole. 
We have considered a thin shell which collapses
at the speed of light as a convenient idealization. 
The geometry is obtained by connecting the flat spacetime inside the shell with
the black-hole geometry outside the shell. 
%We take into account the back reaction from the negative vacuum energy, 
%which is the quantum effect in the energy-momentum tensor in the vacuum. 

We have focused on the spherically symmetric configurations, 
and used the s-wave approximation.
The vacuum energy-momentum tensor is assumed to be approximated by 
that of the 2-dimensional scalar field,
so all angular components vanish. 
%%% 0406-1
\footnote{
This model for the vacuum energy-momentum tensor is widely studied
in the literature for 4D black-hole geometry;
see, e.g. 
\cite{Davies:1976ei,Parentani:1994ij,Brout:1995rd,Ayal:1997ab,Fabbri,Barcelo:2007yk}.
}
%%% 0406-2
The geometry is then uniquely fixed by the initial condition 
in the past null infinity
and the junction condition at 
%YM-12/16
the collapsing shell.
% for the geometry outside it. %the collapsing shell. 
%-12/16
By taking the initial condition such that 
the incoming energy is absent in the asymptotic region, 
the negative vacuum energy appears near the Schwarzschild radius, 
and the areal radius has a local minimum 
slightly larger than the Schwarzschild radius
on each null surface of constant retarded time ($u = \text{const}$).

Assuming that
the angular components of the energy-momentum tensor vanish
on the collapsing null shell,
%it has only (delta functional) incoming energy $T_{vv}$.
the outgoing energy $T_{uu}$ must be continuous across the shell, 
and hence it has to be zero just outside the shell. 
This implies that the Hawking radiation can take no energy directly from the shell. 
Furthermore,
the quantum energy-momentum tensor on top of the shell 
is positive and increasing 
%as the shell is collapsing, and hence, 
such that the total (delta functional) incoming energy on the shell is increasing. 
%This indicates that 
The shell collapses without losing its mass. 
%If the shell with finite mass really shrink to a point, it becomes a singularity. 
%%%-12/23

The presence of the negative energy outside the shell leads to
the structure of Wheeler's bag of gold. 
%YM-12/16
The neck of the bag is nothing but the local minimum of the areal radius. 
It also plays the role of the apparent horizon in the time-dependent case. 
%-12/16
If the neck continues to shrink
and finally closes at some point, 
the interior of the neck is disconnected from the external universe,
and there would be an event horizon. 
However,
the semi-classical Einstein equation 
is only the low-energy effective theory for scales much larger than the Planck length. 
The neck may stop shrinking at a finite size
possibly of the Planck scale. 
We cannot tell within the framework of the low-energy effective theory
whether or not 
there would be an event horizon.

\begin{figure}
\vskip1cm
\begin{center}
\includegraphics[scale=0.4,bb=0 90 330 180]{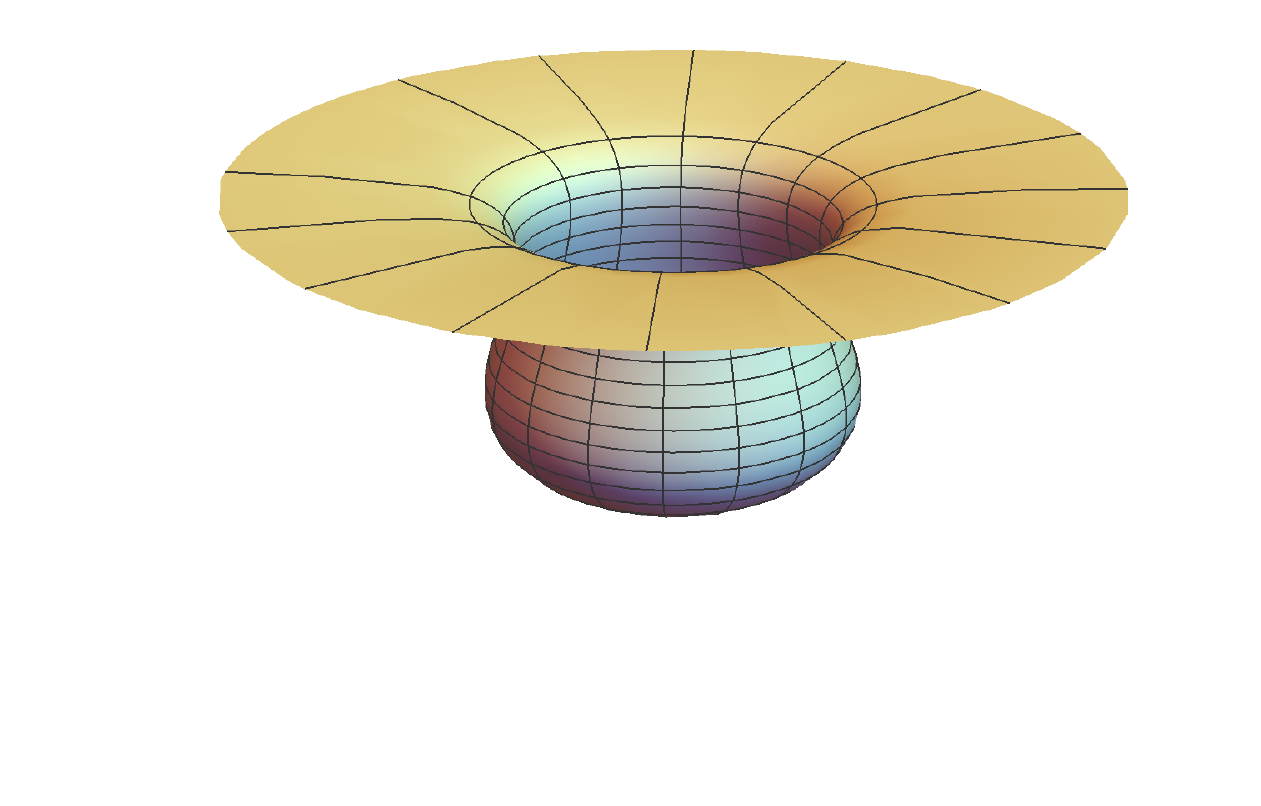}
\includegraphics[scale=0.4,bb=0 90 330 180]{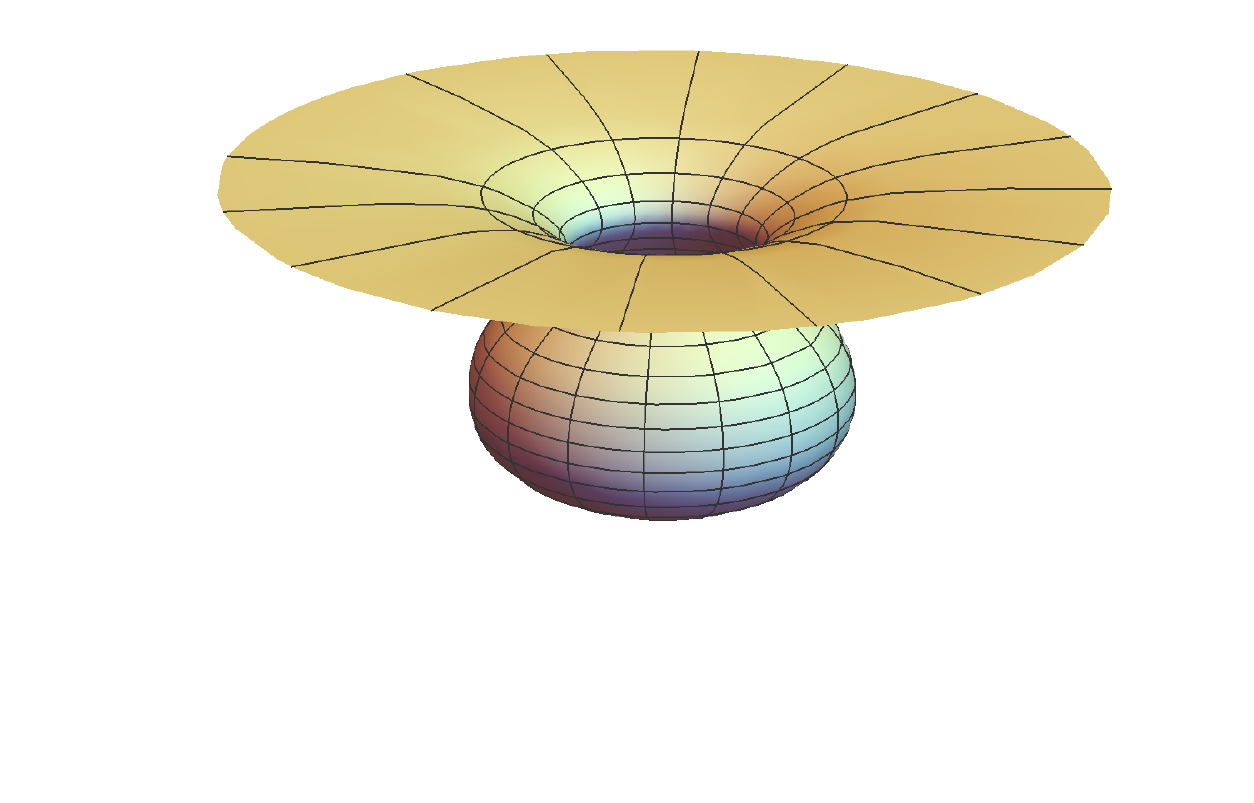}
\includegraphics[scale=0.4,bb=0 90 330 180]{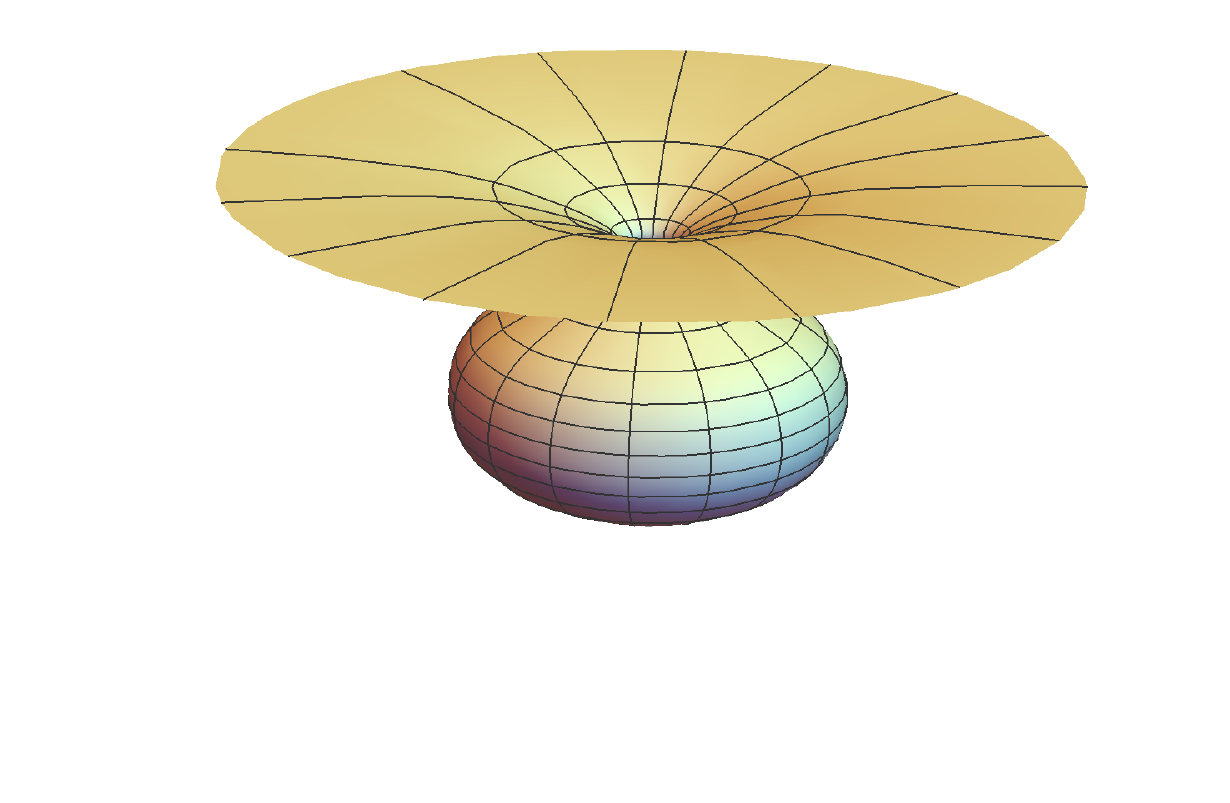}
\caption{\small 
The schematic diagrams for an evaporating black hole 
that ends up with a Planckian neck and a large internal space \cite{Ho:2018jkm}. 
}
\label{fig:WH1}
\end{center}
\end{figure}

\begin{figure}
\vskip1cm
\begin{center}
\includegraphics[scale=0.4,bb=0 90 330 180]{BH-1.pdf}
\includegraphics[scale=0.4,bb=0 65 330 155]{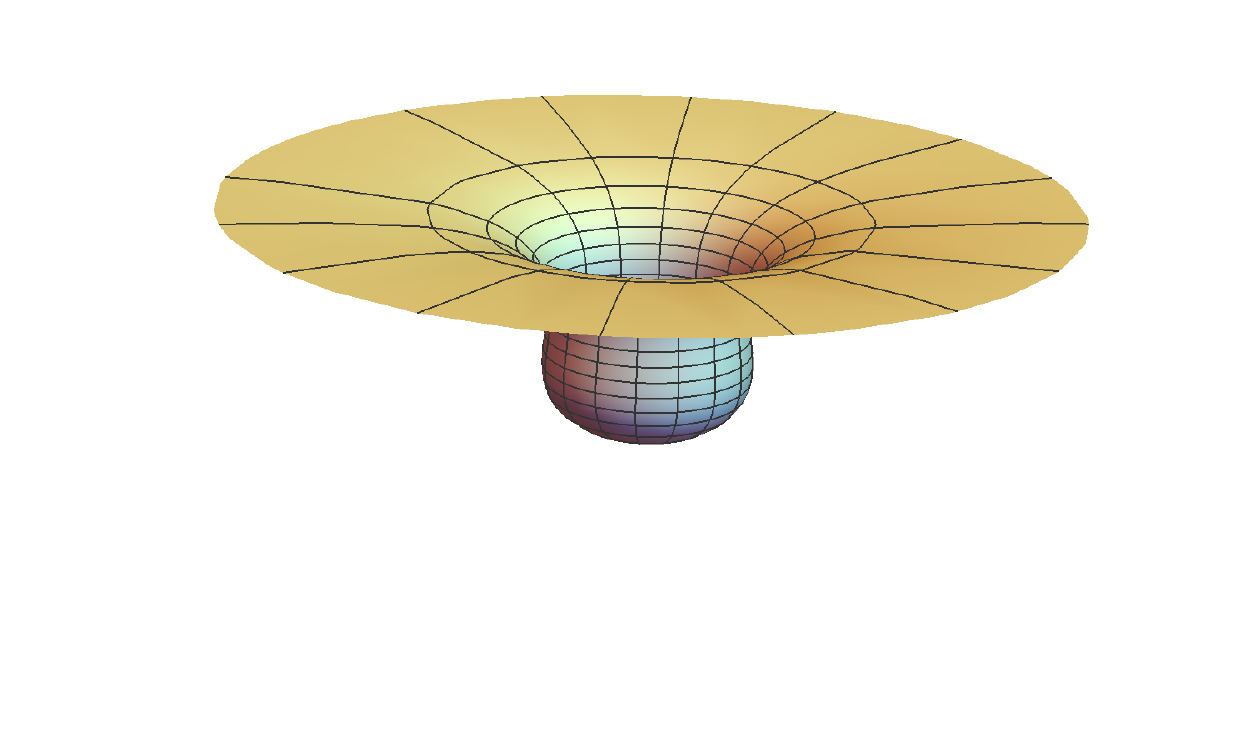}
\includegraphics[scale=0.4,bb=0 -20 330 60]{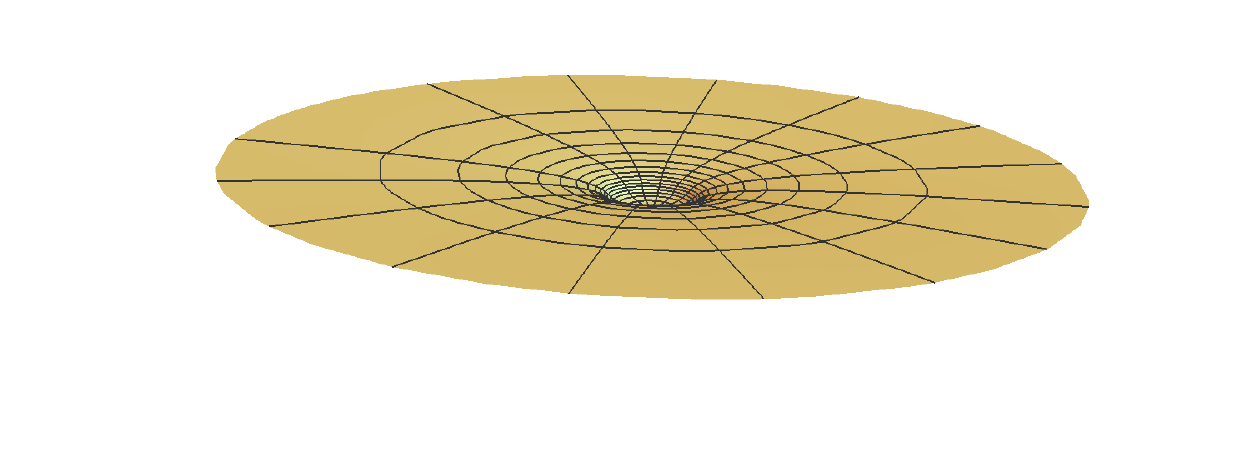}
\caption{\small 
The schematic diagrams for an evaporating black hole
for which the interior space shrinks together with the neck \cite{Ho:2018jkm}.
}
\label{fig:WH2}
\end{center}
\end{figure}

It is interesting to know
whether the scale of the interior space
is still large when the black hole is nearly completely evaporated. 
Our results show that the decreasing rate of the areal radius of the collapsing shell 
becomes nearly zero
while the difference of the areal radii between the shell and the horizon becomes large. 
That is,
the areal radius of the shell 
can become much larger than that at the apparent horizon at the same time
(in terms of the $u$-coordinate).
%YM-4/8
In \cite{Ho:2018jkm}, it was argued that 
there are two possible scenarios depending on whether 
the size of the interior space shrinks together with the neck, 
which are shown in Figs.~\ref{fig:WH1} and \ref{fig:WH2}. 
Our result in this paper implies that in terms of the areal radius, 
the scenario of Fig.~\ref{fig:WH1} is realized for the gravitational collapse of the thin shell. 
%-4/8
This is related to the fact that the energy of the shell cannot be turned into Hawking radiation
in the absence of high-energy events, 
and the shell keeps its initial mass until it can no longer be described
in the low-energy effective theory.
A crucial assumption related to this conclusion is that 
the energy-momentum tensor on the shell does not have 
non-zero angular components. 
If the collapsing shell has tangential pressure in the angular directions, 
the outgoing energy flux can be discontinuous across the shell
without violating energy conservation.
In this case, it is possible that
the outgoing energy appears just outside the shell
and the Hawking radiation can directly take the energy of the shell away,
as the case of the model proposed in Refs.\cite{Kawai:2013mda,KMY}.

As the proper size inside the collapsing shell 
is determined by its areal radius 
since the spacetime is flat inside the shell,
the size of the space inside the apparent horizon 
is much larger than
what is suggested by the size of the neck
when a large portion of the black-hole mass is evaporated. 

Our calculation suggests that
the proper distance between the collapsing shell 
and the apparent horizon is as small as the Planck length.
\footnote{
A concise proof of the Planckian proper distance
between the collapsing shell and the trapping horizon
for a wider class of models will be given elsewhere
\cite{Ho:2019qiu}.
}
The positive energy of the shell and the negative vacuum energy 
are placed within this small region of the Planck scale. 
In particular, most of the vacuum negative energy is placed 
in the deeper region whose size in proper length is much smaller than the Planck scale. 
Since the semi-classical picture of the spacetime is not good for such a small region, 
the positive and negative energy in this small region should be course-grained 
as a more appropriate picture of the spacetime in the low-energy effective theory. 

Strictly speaking,
the Planck-scale separation between the apparent horizon and the shell
means that
we cannot really distinguish the location of the collapsing shell from that of the neck
in the context of a low-energy effective theory
with a cutoff length scale larger than the Planck length.
Whether the collapsing matter has entered the apparent horizon is 
thus a question beyond the scope of low-energy effective theories. 
We can only conclude that the apparent horizon 
and the collapsing shell are approximately around the same place. 
It is hence interesting to note that,
in the model proposed in Refs.\cite{Kawai:2013mda,KMY}
(with a different assumption about the vacuum energy-momentum tensor),
the collapsing matter stays outside the apparent horizon by a Planckian distance,
so that in fact the apparent horizon does not emerge.
The difference between models with and without apparent horizons 
might be subtler than what people have expected.

\subsubsection*{Acknowledgments}

The author would like to thank Hikaru Kawai, Piljin Yi, Tatsuma Nishioka for discussions.
P.M.H. thanks iTHEMS at RIKEN,
University of Tokyo and Kyoto University
for their hospitality during his visits where a significant part of this work was done.
P.M.H.\ is supported in part by the Ministry of Science and Technology, R.O.C. 
and by National Taiwan University. 
The work of Y.M.\ is supported in part by JSPS KAKENHI Grants No.~JP17H06462.
Y.Y.\ is partially supported by Japan Society of Promotion of Science (JSPS), 
Grants-in-Aid for Scientific Research (KAKENHI) Grants No.\ 18K13550 and 17H01148. 
Y.Y.\ is also partially supported by RIKEN iTHEMS Program.

\end{document}